\documentclass[12pt,preprint]{aastex}

 \newcommand{\kms}{km s$^{-1}$}
 
 \newcommand{\HII}{\ion{H}{2}}

 \newcommand{\moy}{M$_\odot$~yr$^{-1}$}

\shorttitle{Cygnus-X Bowshocks}

\begin{document}

\slugcomment{2009 December 7}

\title{OB Stars \& Stellar Bowshocks in Cygnus-X:
A Novel Laboratory Estimating Stellar Mass Loss Rates}

\author{Henry A. Kobulnicky}
\affil{Department of Physics \& Astronomy \\  1000 E. University \\
University of Wyoming \\ Laramie, WY 82071 
\\ Electronic Mail: chipk@uwyo.edu}

\author{Ian J. Gilbert\altaffilmark{1} }
\affil{Department of Physics \\ 100 Campus Drive \\ Grove City College 
\\ Grove City, PA 16127
\\ Electronic Mail: gilbertij1@gcc.edu}

\altaffiltext{1}{Also at Department of Physics \& Astronomy \\ 1000 E. University \\ 
University of Wyoming \\ Laramie, WY 82070}

\author{Daniel C. Kiminki}
\affil{Department of Physics \& Astronomy \\ 1000 E. University \\ 
University of Wyoming \\ Laramie, WY 82070
\\ Electronic Mail: kiminki@uwyo.edu}


\begin{abstract}

We use mid-IR images from the {\it Spitzer Space Telescope} Cygnus~X
Legacy Survey to search for stellar bowshocks, a signature of early
type ``runaway'' stars with high space velocities.  We identify ten
arc-shaped nebulae containing centrally located stars as candidate
bowshocks.  New spectroscopic observations of five stars show that all
are late O to early B dwarfs, while one is a previously classified B0.2
giant.  These stars have moderate radial velocities, differing by
${\Delta}V<$10 \kms\ from members of the Cygnus~OB2 Association.  The
spectral energy distributions of the other four stars are consistent
with late O to early B dwarfs at the nominal $\sim$1.6 kpc distance of
Cyg~OB2.  Our morphologically selected sample of bowshock candidates
encompasses diverse physical phenomena.  Three of the stars appear to
be pre-main-sequence objects on the basis of rising SEDs in the
mid-IR, and their nebulae may be photon-dominated regions (PDRs)
illuminated by the central star but shaped by external sources such as
winds from Cyg~OB2.  Four objects have ambiguous classification.
These may be partial dust shells or bubbles. We conclude that three of
the objects are probable bowshocks, based on their morphological
similarity to analytic prescriptions.  Their nebular morphologies
reveal no systematic pattern of orientations that might indicate
either a population of stars ejected from or large-scale
hydrodynamic outflows from Cyg~OB2.  The fraction of runaways among OB
stars near Cyg~OB2 identified either by radial velocity or bowshock
techniques is $\sim$0.5\%, much smaller than the $\sim$8\% estimated
among field OB stars. We discuss possible reasons for this difference.   
We also obtained a
heliocentric radial velocity for the previously known bowshock star,
BD+43\degr3654, of -66.2 $\pm$ 9.4 \kms, solidifying its runaway
status and implying a space velocity of 77$\pm$10 \kms.  We use the
principles of momentum-driven bowshocks in conjunction with the
observed sizes, bowshock luminosities \& spectral energy
distributions, and dust/PAH emission models to arrive at a novel
method for estimating stellar mass loss rates.  Derived mass loss
rates range between 10$^{-7}$ and few$\times10^{-6}$ \moy\ for the
three O5V -- $\sim$B2V stars identified as generating bowshocks. These
values are at the upper range of, but broadly consistent
with, estimates from other methods.  We calculate a
relatively large mass loss rate of 160$\times10^{-6}$ \moy\ for O4If star
BD+43\degr3654 using the same method.

\end{abstract}


\section{Introduction } 

Most OB stars lie within the associations in which they were
born. However, a significant fraction (17\%--50\%) having high
(30--200 \kms) space velocities wander far from such associations
and are known as ''runaways'' \citep{blaauw, stone82, conti77,
gies86}.  Often, these stars' proper motions allow them to be
traced back to a known OB association in which they presumably formed.
The speed necessary for a star to be classified as a runaway varies
from 30--40 \kms\ depending on author. We adopt the 30 \kms\ criterion
used by \citet{gies86}.

\citet{blaauw} suggested that runaway stars are ejected from OB
associations when one member of a binary system sheds a large
fraction of its mass, as in a type II supernova. In this case (known
as the binary-supernova scenario), the surviving companion is released
with a velocity comparable to its previous
orbital velocity. \citet{blaauw} cited the low binary frequency of
runaway stars as evidence for this model, though \citet{gies86} found a
few runaway binaries in their survey.   Whether and how frequently 
supernovae in binary systems produce an unbound, high-velocity 
O star is an active line of investigation.

\citet{gies86} proposed that runaway stars are ejected during
close encounters between binary systems in OB associations. This
model, known as the dynamical-ejection scenario, explains the
existence of binary runaway systems, since hard encounters among three
or more stars can produce both single runaway stars and binary runaway
systems.  \citet{leonard88, leonard90} performed numerical simulations
of young star clusters, showing that dynamical encounters can readily
account for the number of observed runaways and that runaway binary
systems are possible.

\citet{hoogerwerf} used precision proper-motion measurements of nearby
runaways to find evidence for both binary-supernova and dynamical
ejection mechanisms.  They concluded that the runaway $\zeta$ Oph
likely resulted from the dissociation of a binary system wherein the
other member became the pulsar PSR J1932+1059.  They also concluded 
that the runaways AE Aurigae and $\mu$ Columbae and the binary system
$\iota$ Orionis were produced in a binary-binary encounter.

Traditionally, runaway stars reveal themselves through their high proper
motions or unusually large radial velocities \citep{gies86, mdzin}.
Identification using radial velocities requires high-precision
spectra, and this technique is sensitive only to runaways with large
projected radial velocity components.  Identification on the basis of
high proper motions requires precise astrometric data, such as those
made by the Hipparcos \citep{perryman} mission, along with secure
distance measurements.  This technique is only sensitive to runaways
with large tangential motions.

Runaway O stars have also been identified by detecting the tell-tale
``bowshocks'' produced when the stellar winds of stars traveling
supersonically impinge upon the surrounding interstellar medium.
Bowshocks appear as symmetric arc-shaped features with apsides oriented
in the direction of the stars' motions and at distances from the star
determined by momentum balance between the wind and the ambient
medium.  \citet{vanb88}, \citet{vanb95}, and \citet{noriega97} used
Infrared Astronomical Satellite (IRAS) images to locate arc-shaped
features associated with high-velocity O stars.  These studies noted
that a variety of phenomena can produce bowshock-like morphologies,
including partial stellar wind bubbles and dust shells, \ion{H}{2} regions
with density gradients, and genuine high-velocity stars.  Indeed,
we conclude that our sample is comprised of a mixture of such objects.   
Several more recent studies have used H$\alpha$ surveys, Midcourse Space
Experiment (MSX) and {\it Spitzer Space Telescope (SST)} infrared
images to detect bowshocks and associate them with high-velocity O
stars \citep{brown05, comeron07, bomans08, povich}.  Bowshocks are also
seen in environments such as the Orion Nebula
where winds and jets from young encounter the O-star winds that power
the nebula (e.g., the $HST$ images of \citet{bally00}).    

In this paper we use infrared images from the {\it SST} Cygnus-X
Legacy Survey \citep{hora} to identify bowshock candidates in the
vicinity of one of the Galaxy's richest OB associations.  The
association Cygnus~OB2, at the heart of the Cygnus~X region, contains
over 100 OB stars, including an O3If and O4If \citep{mt91, hanson03,
comeron02, comeron07}.  The spectroscopic OB survey of 142 stars in Cygnus OB2 by
\citet{kiminki07} did not detect any runaway stars on the basis of
radial velocities.  However, in a region as rich in young massive
stars as Cygnus~OB2, one would also expect to find stars ejected with
large motions in the plane of the sky.  \citet{comeron07} implicate
Cygnus~OB2 as the origin of one such star, the O4If runaway
BD+43\degr3654, using the morphology of the bowshock seen in {\it MSX}
images.  Given the superior sensitivity and angular resolution of {\it
SST}, a more complete search may now be conducted for runaway stars
hosting bowshocks in this region.  Herein, we describe the discovery of
ten candidate bowshocks near Cygnus~OB2, and we report followup
optical spectroscopy that allows us to determine the spectral types
and radial velocities for a subset of the central stars that power
each candidate.  We refer to the bowshock candidates and their central
stars separately as Bowshocks (BS) 1--10 and Stars 1--10.  We also
investigate a novel method of determining the mass loss rates of these
stars using the principles of momentum-driven bowshocks.

Modern distance estimates to Cyg~OB2 range from 1.7 kpc \citep{mt91,
torres} to 1.4 kpc \citep{hanson03}.  We adopt d=1.6 kpc, at which
distance 1 pc corresponds to 130\arcsec\ on the sky.

\section{Data}

\subsection{Spitzer Space Telescope}

We retrieved data from the {\it SST} archive obtained as part of the
{\it Spitzer} Legacy Survey of the Cygnus-X Complex \citep{hora} using
data from the Infrared Array Camera (IRAC) at 3.6, 4.5, 5.8, and 8.0
microns \citep{fa04} and the Multi-band Infrared Photometer for
Spitzer (MIPS) bandpasses at 24 \& 70 microns \citep{rieke}.  One of
us (HAK) visually examined mosaiced single-band and multi-color images
of the $\sim$24 square degree survey area to identify bowshock
candidates on the basis of symmetric arc-shaped structures less than a
few arcminutes in size enclosing a symmetrically placed stellar
source.  The complex ISM structure in this region, coupled with the
large dynamic range of features, renders bowshock identification a
subjective and imprecise task.  In practice, the 8-micron band
(sensitive to emission from PAH features excited by non-ionizing
photons) and the 24-micron band (sensitive to emission from warm dust)
were the principal images used to locate candidate bowshocks.  While a
plethora of arc-like structures appear in both bands, very few
exhibited a high degree of symmetry and included a point source near
the apsis.  In total, ten objects were deemed to be bowshock
candidates and were retained for further investigation.
Table~\ref{star.tab} lists the equatorial \& Galactic coordinates and
2MASS JHK \& $SST$ IRAC photometry of the central stars for each
object.  The final column provides common stellar cross
identifications from other works, where applicable.
 
\begin{figure}
\plotone{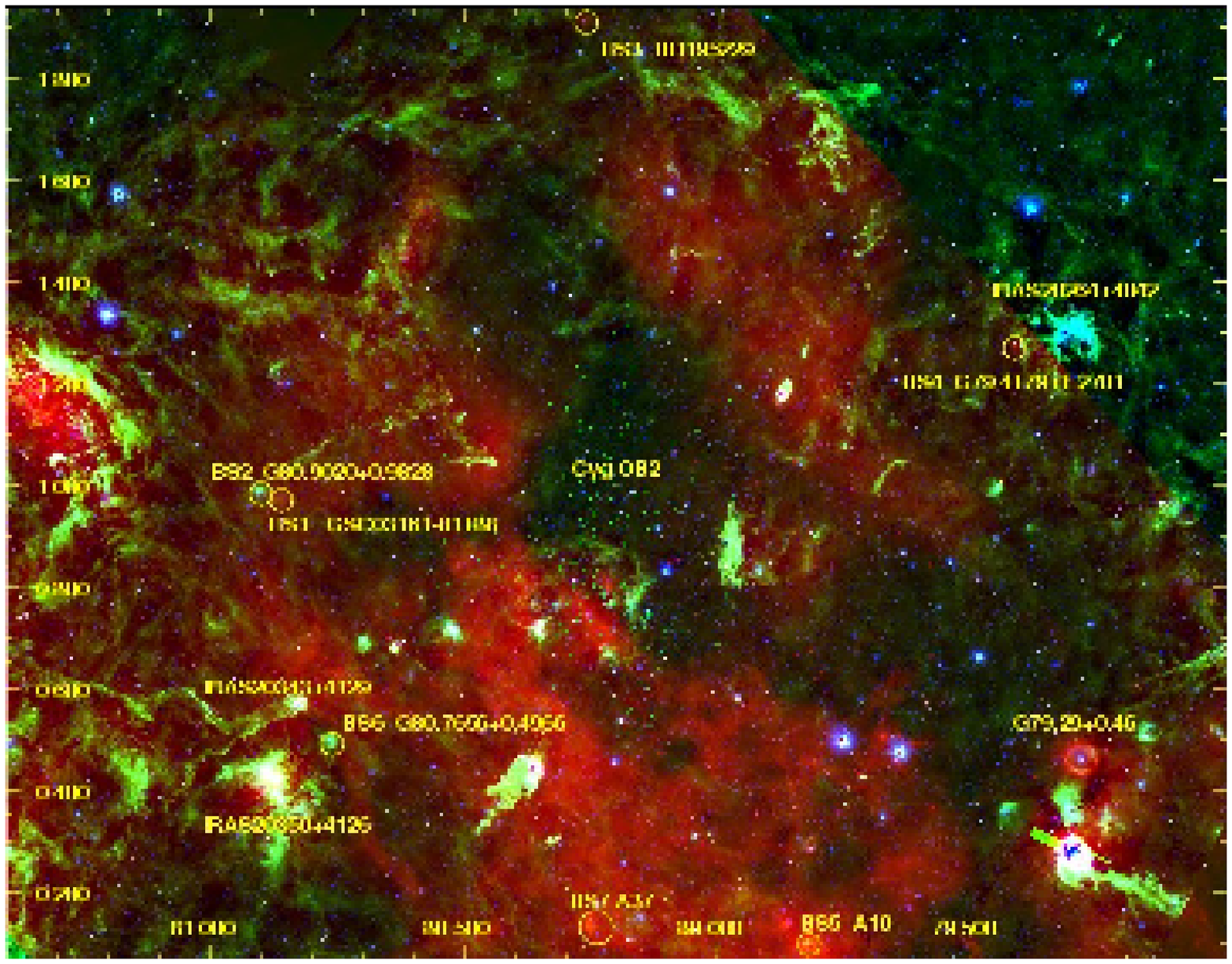}
\caption{{\it Spitzer} three-color image in Galactic coordinates of the Cygnus-X region centered on Cygnus~OB2 with
[4.5] in blue, [8.0] in green, and [24] in red.  
Seven of the ten bowshock candidates are labeled with their bowshock number
and identifications for the central stars from Table~\ref{star.tab}.
Several other prominent IRAS infrared sources are labeled, some of which are the heads 
of gaseous pillars pointing toward Cyg~OB2, such as IRAS20343+4129.      
\label{overview} }
\end{figure}

Figure~\ref{overview} is a 3-color image $\sim2$\degr\ square centered
on Cygnus OB2 ($\ell=80.3$\degr, $b=+1.0$\degr) showing the location
of the first seven candidate bowshocks.  Blue, green, and red
represent the 4.5, 8.0, and 24 micron images, respectively.  Objects
8--10 have Galactic latitudes $b<0.15$\degr\ and lie off the lower
edge of Figure~\ref{overview}.  Figures~\ref{BS1+2} through \ref{BS10}
show enlarged views of each bowshock candidate using the same color
scheme as Figure~\ref{overview}.  We discuss each object in detail in
\S 3.

\subsection{WIRO}

We obtained optical spectra for several of the stellar sources located
near the apsides of candidate bowshocks.  We observed five of the
stars (Stars 1, 2, 3, 5, and 8) at the Wyoming Infrared
Observatory's (WIRO) 2.3 m telescope equipped with the Longslit
spectrograph on 2008 June 23--30, 2009 May 20, and 2009 September 16.  The
1800~l~mm$^{-1}$ grating was used in first order. The spectral
coverage was 5210--6680 \AA. Exposure time varied from 360~s for Star
3 (HD~195229) to 3600~s for Stars 2 and 5, yielding signal-to-noise
ratios ranging from 50:1 for Star 5 to 450:1 for Star 2. Copper-Argon
lamp exposures were taken after each stellar spectrum for wavelength
calibration.

The spectra were reduced, extracted, wavelength calibrated, and
Doppler corrected to the heliocentric frame using standard
IRAF\footnote{IRAF is distributed by the National Optical Astronomy
Observatories, which are operated by the Association of Universities
for Research in Astronomy, Inc., under cooperative agreement with the
National Science Foundation.} tasks in the KPNOSLIT package. The
absolute radial velocities were checked against the standards HD161096, HD
161797, HD131156, HD171391, and HD188512.  
Table~\ref{star.tab} lists the heliocentric radial velocities, which are
typically accurate to 6 \kms\ rms.

\subsection{WIYN}
We also obtained spectra of BD+43\degr3654 and HD195229 during an
observing run at the WIYN\footnote{The WIYN Observatory is a joint
facility of the University of Wisconsin-Madison, Indiana University,
Yale University, and the National Optical Astronomy Observatories.}
3.5 m telescope on 2008 June 11--16 using the Hydra
multispectrograph. We used the Red camera and 3$''$ blue fiber cables
with the 1200~l~mm$^{-1}$ grating in second order.  Exposure times
were 600~s and the spectral coverage was 3820--4500 \AA. Copper-Argon
lamp exposures were taken to wavelength calibrate the
spectra.  The spectra were reduced, extracted, wavelength calibrated, and
Doppler corrected to the heliocentric frame using standard IRAF tasks
in the TWODSPEC and RVCOR packages. Radial velocities of standards HD
131156, HD146233, HD161096, HD161797, and HD171391 from this run
agreed with published values to $\pm$2 \kms.

\section{Analysis}
\subsection{Bowshock Identification and Images}

Figure~\ref{overview} shows a portion of the Cygnus-X region
surrounding its most massive and energetic constituent, Cygnus~OB2.
This region encompasses a complex array of nebulae and star-forming
regions located at various distances along the line of sight, with
most of the activity located in the local or ``Orion'' spiral arm at a
distance of 1--2~kpc.  \citet{odenwald} present a schematic
3-dimensional representation of the Cygnus-X region, while
\citet{schneider} provide velocity-resolved molecular maps of Cygnus-X
and individual star-forming objects nearby.  One apparent feature of
this region is the relative absence of dust (red) and PAH emission
(green) at the location of Cyg~OB2 itself, except at the southern
edge.  This apparent ``hole'' is consistent with a cavity cleared by
the winds of $>$100 evolving massive stars.  Some of the prominent
star-forming regions labeled in Figure~\ref{overview} clearly
constitute the heads of gaseous pillars, most of which point toward
Cyg~OB2, the dominant source of luminous and mechanical energy in this
vicinity.  These pillars are reminiscent of those in M~16
\citep{indebetouw07, hester96}, RCS~49 \citep{whitney04} and other
massive star-forming regions.

The seven bowshock candidates labeled in Figure~\ref{overview} lie
around the periphery of Cyg OB2, none falling within the main
concentration of Cyg~OB2 stars but all within $\sim$5--8 pc of the
canonical Cyg~OB2 center.  The orientations of the features, seen more
clearly in Figures~\ref{BS1+2} through \ref{BS10}, appear random,
having no preferential alignment with respect to Cyg~OB2.
Generically, bowshock orientation reveals the relative motion between
the star and surrounding matter.  We might have expected, if the stars
were all runaways from Cyg~OB2, that the bowshocks would point away
from the Association in the direction of the stars' motions.  On the
other hand, if the hydrodynamics of material surrounding Cyg OB2 were
dominated by a hot outflow from an over-pressured region, we might
have expected many of the bowshocks to point back toward the source of
mechanical energy, as do three bowshocks near M~17 and at least one
near RCW~49 \citep{povich}.  The absence of either signature suggests
that 1) the central stars driving the bowshocks, if they are runaways,
have different origins, and 2) there is no large-scale supersonic wind
emanating from Cyg~OB2.  Only candidates 2 and 4 have orientations
consistent with their stars being runaways having an origin within
Cyg~OB2.

\subsection{Nebular and Stellar SEDs}

The colors of the candidate bowshock nebulae , best seen in the
individual zoomed images in Figures~\ref{BS1+2} through \ref{BS10}, at
first glance reveal two classes of objects.  Candidates 1,3,4,7 are
seen exclusively at 24 microns and longer wavelengths while the rest
exhibit strong emission in the IRAC bandpasses with some additional
contribution at longer wavelengths.

We performed aperture photometry of each bowshock candidate in each of
the six bandpasses using the {\it SSC} post-basic calibrated data
mosaic images by assigning irregular crescent-shaped polygonal
apertures defined visually on the [8.0] and/or [24] images.  We also
manually defined background regions surrounding each bowshock.
Because these objects lie in complex regions of diffuse emission, care
was taken to select multiple background regions judged to be
representative of the local background away from the nebula but
within several arcminutes.  We applied aperture corrections, as
recommended in the IRAC and MIPS calibration web
pages\footnote{http://ssc.spitzer.caltech.edu/irac/calib/ ;
http://ssc.spitzer.caltech.edu/mips}, using values appropriate to a
circular aperture of equivalent area to the polygonal apertures.
These corrections are less than 1.3 in all cases and are sometimes
less than unity for the IRAC bands.  Given the imprecision of the
aperture corrections and the high and variable background levels, we
adopt a minimum uncertainty of 20\% in all bands.
Table~\ref{phot.tab} gives background-subtracted aperture fluxes in
mJy for all of the  nebulae.  In some cases, only 1$\sigma$
upper limits are given, mostly in the IRAC bands where some objects
are not detected.

For each bowshock candidate we were able to identify a symmetrically placed
point source, presumed to be the energizing star, behind the apsis.
Figure~\ref{ccd} is a 2MASS JHK color-color diagram showing each of
the ten stars ({\it diamonds}) associated with the bowshocks.  The
solid and dashed curves ({\it asterisks and crosses, respectively})
show a fiducial main sequence and supergiant sequence.  A solid line
illustrates the reddening vector for $A_V=5$ mag.  The majority of
stars lie in a close group near H-K=0.3, J-H=0.6, consistent with
early-type stars seen behind 4--7 magnitudes of visual extinction.
This range is similar to other OB stars in Cyg~OB2 \citep{mt91},
suggesting that they lie at a similar distance.  Stars 4 and 6 are
much redder, suggesting 12--15 magnitudes of visual extinction, under
the assumption that these are also early-type stars.  The K-band
magnitudes for these two stars are 2--3 magnitudes fainter than the
other targets, indicating either a larger distance or the effect of
localized regions of high extinction.  Star 6, in
particular, is surrounded by a more substantial region of warm dust
and photo-excited material, consistent with localized extinction.
Star 3 (HD195229), a known B0.2III, is much less red, consistent with
$\sim$1 magnitude of visual extinction, and consistent with the known
variability of extinction across Cyg~OB2 \citep{mt91}.

We performed aperture photometry on the central star of each bowshock
at the mid-IR IRAC bandpasses using the {\it SSC} basic calibrated
data frames, a 10-pixel circular aperture for which the aperture
correction is minimal, and a much larger annulus to measure the
diffuse background levels.  Most of the stars were saturated at IRAC
[3.6] and [4.5] bands in the 10.4~s exposures, so the
high-dynamic-range (HDR) 0.4~s exposures were used.
Table~\ref{star.tab} lists these fluxes and their uncertainties.

Figure~\ref{BSsed} shows the spectral energy distributions of the
bowshock candidates and central stars at the 2MASS, IRAC and two MIPS
bandpasses.  Asterisks denote photometry of the bowshock alone, while
diamonds denote the star plus bowshock.  The bowshocks are not
detected at 2MASS JHK bands and are somtimes undetected in the IRAC
bandpasses.  In only three cases (stars 6, 8, and 9) are pointlike
sources detected at [24], indicative of dust in the immediate vicinity
of the stars.  Uncertainties are always smaller than the plotted
points.  The solid curves are blackbody fits to the JHK photometry,
while the dotted curves are fits to the IRAC and MIPS data using
scaled \citet{draineli07} dust emission models.  In Sections 3.4 and 4
we discuss the details of the fitting, the models, and model results
for each object individually.

\subsection{Bowshock Candidate Morphologies}

The shape of a momentum-driven bowshock is a universal function that
scales as the ``standoff distance'', $R_0$, between the star and the
bowshock apsis, as shown analytically by \citet{baranov, dyson,
vanb90, wilkin96}.  Using the formulation of \citet{wilkin96} Equation
9, we generated synthetic bowshock ``images'', applying arbitrary
size, rotation, and intensity scale factors to facilitate comparison
with the data.  Figure~\ref{sim} shows a selected subset of bowshock candidates
(1, 2, 3, 5, 6, and 7) at [8.0] ({\it green}), [24] ({\it red}) and
the scaled synthetic bowshock ({\it blue}).  The simulated bowshocks
are all shown at zero degrees inclination, while the inclinations of
the bowshock candidates are unknown.

The morphologies of candidates 5 and 7 (central stars O9V and O5V)
show striking similarity to the theoretical shape. The agreement is
best at [24] where the bowshock can be traced to larger radial
distances than at [8.0] where the bowshock is less extended and
exhibits a more irregular morphology.  For object 2, there is also
good agreement, although the bowshock itself is faint and difficult to
see because of the large dynamic range in the images.  For objects 1,
3, and 6 the morphological agreement is considerably less good.  We
take this as evidence for a physical origin other than a bowshock.
The morphologies of the other objects 4, 8, 9, and 10 (not pictured) also differ
from the canonical bowshock shape, appearing more irregular and
clumpy.  We interpret the good agreement between the model shape and
objects 2, 5, and 7 as support for the momentum-driven bowshock nature
of these objects.

\subsection{Individual Objects}
\subsubsection{Bowshock Candidate 1 and Star 1 (GSC03161-01188)}

Bowshock candidate 1, pictured in Figure~\ref{BS1+2}, is associated
with the V=13.3 star GSC03161-01188 from the Guide Star Catalog, also
known as \#1288 from the work of \citet{rlp}.  Figure~\ref{spec1}
(bottom) shows the WIRO spectrum of GSC03161-01188.  We classify
GSC03161-01188 as O9V on the basis of the He{\sc I}/He{\sc II} line
ratios and by comparison to a \citet{jacoby} spectral library O9V star
(top).  We compute a spectrophotometric distance for the K=8.21 star
GSC03161-01188 by adopting a visual extinction of 4.5 magnitudes from
Figure~\ref{ccd} ($A_K=0.51$ mag), an intrinsic color V-K=$-$0.85, and
assuming an absolute magnitude for an O9V star of $M_V=4.05\pm0.2$
\citep{martins05}.  This yields a spectrophotometric distance of 1500
$\pm$ 150 pc for GSC03161-01188, consistent with the adopted distance
to Cyg~OB2, $\sim1.6$~kpc \citep{hanson03,mt91,schulte}.

Cross correlating the spectrum of GSC03161-01188 with a synthetic
model atmosphere O9V spectrum from \citet{lanz}, as described more
fully in \citet{kiminki07}, yields a heliocentric radial velocity of
$-17\pm6$ \kms.  This velocity is consistent with, but more negative
than, the average radial velocities of Cyg~OB2 members, $-10\pm6$
\kms\ surveyed by \citet{kiminki07}.  


The spectral energy distribution in Figure~\ref{BSsed} shows that
this nebula is detected only at [24] and [70].  Stellar flux
dominates the 2MASS and IRAC bandpasses.  The solid line shows
the SED of an O9.5V star, approximated as a blackbody with
$T_{eff}$=30,000~K and R=8 $R_\odot$, \citep{martins05} which, at a
distance of 1.6 kpc with $A_V=4.5$ mag, provides a good fit to the
near-IR data.  This is consistent with our spectral classification
based on Figure~\ref{spec1}.

Given the lack of similarity to the theoretical shape and the modest
radial velocity of the central star, this nebula is unlikely to be a
bowshock.  We assign this object an ambiguous classification.
Possibilities include a dusty asymmetric bubble or a partial shell.

\subsubsection{Bowshock Candidate 2 and Star 2 (G80.9020+0.9828)}

\begin{figure}
\plotone{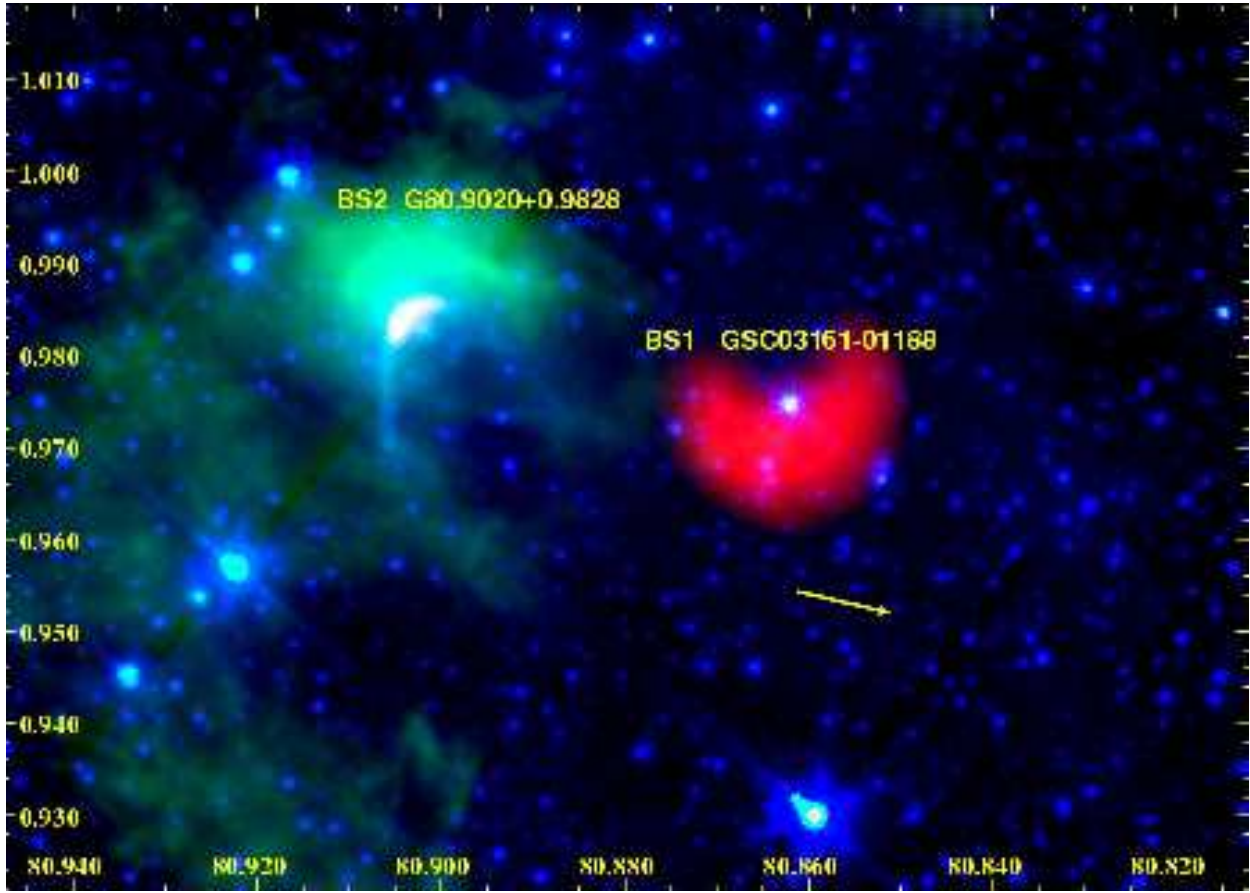}
\caption{{\it Spitzer} three-color image of Bowshock Candidates 1 (GSC03161-01188) and
2 (G80.9020+0.9828) with the same color scheme as Figure~\ref{overview}.  
These two objects present a striking contrast as Bowshock 1 is seen only at 24 microns
and longer wavelengths while Bowshock 2 has strong PAH emission in the IRAC bands and
is weaker at longer wavelengths.  The arrow shows the direction toward the core of Cyg~OB2.  
\label{BS1+2} }
\end{figure}

Figure~\ref{BS1+2} shows bowshock candidate 2 and Star 2
(G80.9020+0.9828).  Figure~\ref{spec2} (bottom) shows the WIRO
spectrum of this star and a comparison B2 dwarf equivalent from the
model atmospheres of \citet{lanz} (top).  The relative strengths of
the \ion{H}{1} 6563 \AA\ and the \ion{He}{1} 6678 \AA\ and \ion{He}{1}
5876 \AA\ lines lead us to classify this star as B2V, plus or minus
several spectral subtypes, owing to the low signal-to-noise spectrum.
Its heliocentric radial velocity is $-12\pm15$ \kms, consistent with
the range of radial velocities observed for Cyg~OB2 members
\citet{kiminki07}.  We compute a spectrophotometric distance from the
2MASS photometry (K=10.5) by adopting a visual extinction
\citep{cardelli89} of 4.5 magnitudes ($A_K=0.51$), an absolute
magnitude for an B2V star of $M_V=-2.4\pm0.4$, an intrinsic V-K color
of -0.4.  This yields a spectrophotometric distance of 2500 $\pm$ 500
pc, where the error is driven primarily by the uncertainty on the
spectral type/absolute magnitude.  A high-surface-brightness apsis
lies 9\arcsec\ (0.07 pc projected separation) from the star toward the
upper left in Figure~\ref{BS1+2}, and there is also a more extended
region of 8 micron emission preceding the apsis to the upper left.
The [24] lima-bean shaped nebulosity and the arc-like structure seen
at [8.0] do not share a common symmetry axis, so it is difficult to
ascertain which wavelength best traces the putative bowshock.
   
Figure~\ref{BSsed} shows that Bowshock 2 is a strong detection at the
IRAC and MIPS bandpasses. The relative faintness of the bowshock in
[4.5] relative to the other three IRAC bandpasses is consistent with
strong PAH emission in the shock. The central star dominates the 2MASS
bandpasses.  Star 2 is fainter than Star 1, and is well fit by a
B2V--B3V star ($T_{eff}$=20,000~K, R=4 $R_\odot$) at a distance of 1.6
kpc with an extinction of $A_V=4.5$ mag ({\it solid line}).

Although this candidate bowshock nebula is faint and difficult to
represent owing to a large dynamic range in the images, the
morphological agreement with the theoretical shape is reasonably good.
The apsis of this nebula is bright at [8.0] and [24] and may indicate
an interaction or illumination of a of a nearby cloud.  We retain this
object as a probable showshock.

\subsubsection{Bowshock Candidate 3 and Star 3 (HD195229)}

Bowshock candidate 3 appears to be generated by its attendant star,
HD195229, listed as B0.2III in SIMBAD.  Figures~\ref{spec3} and
\ref{spec4} show our WIRO and WIYN spectra of this star, from which we
confirm a B0.2III spectral type.  From these spectra we find a
heliocentric radial velocity of $-3\pm2$ \kms, slightly less negative
than other massive stars near Cyg~OB2.  The luminosity for a B0.2
giant is somewhat uncertain because the evolutionary tracks for
massive giant and supergiant stars are nearly vertical in an H-R
diagram \citep{ma08}.  Assuming $M_V=-5\pm0.5$, V-K=-0.8 and $A_V=1.0$
from Figure~\ref{ccd} for this $K=7.3$ star leads to a
spectrophotometric distance of 1800 $\pm$ 500 pc, consistent with the
distances of massive stars in Cyg~OB2.

\begin{figure}
\plotone{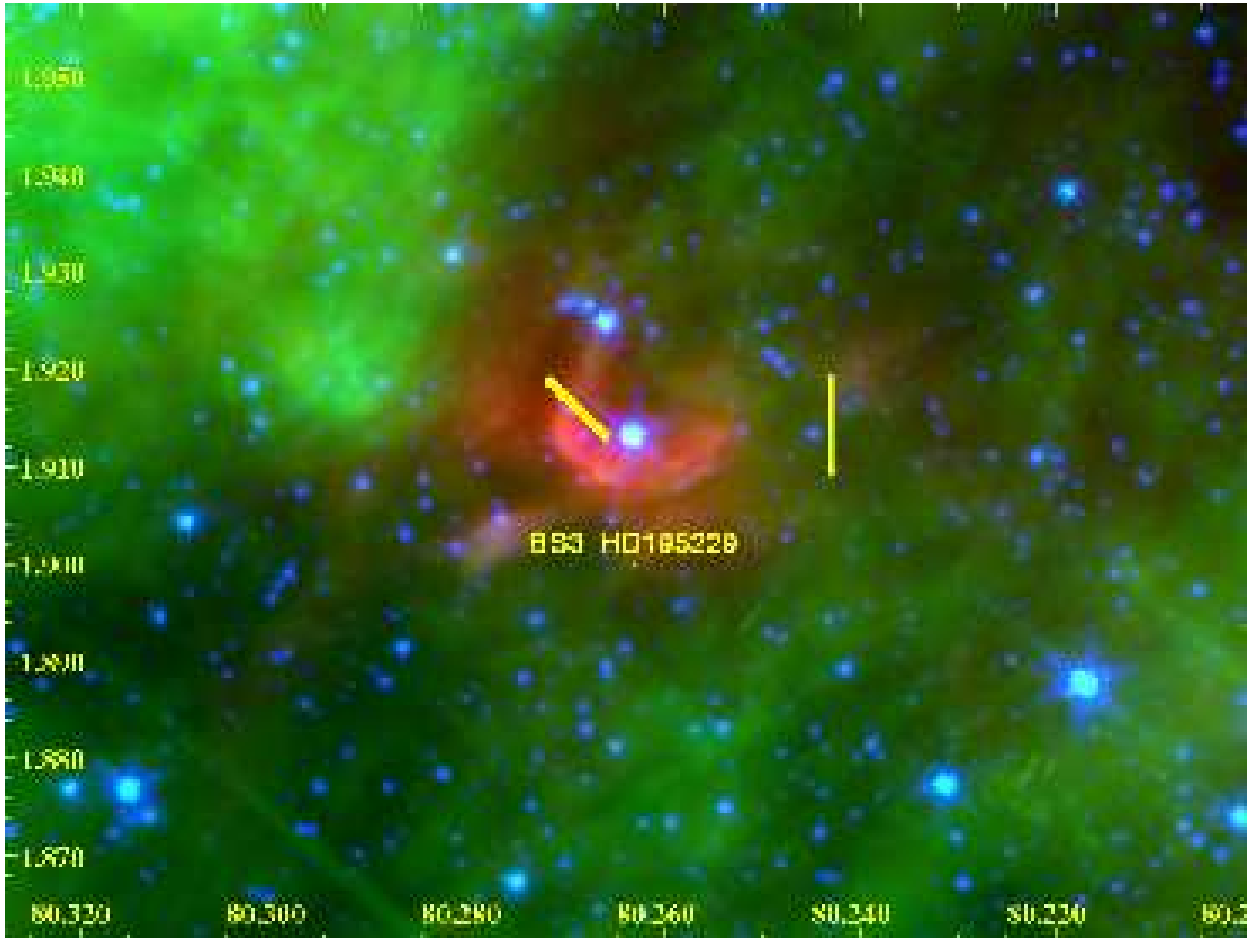}
\caption{Three-color image as in Figure~\ref{overview} 
showing the bowshock candidate around Star 3 (HD195229). The yellow line
indicates HD195229's proper {\it Hipparcos} motion, indicating that there
is a misalignment between the velocity vector inferred from proper motion data and the 
vector inferred from the putative bowshock morphology.  The uncertainty on the position angle of the proper motion
is $\sim$15\degr.
\label{BS3} }
\end{figure}

\begin{figure}
\plotone{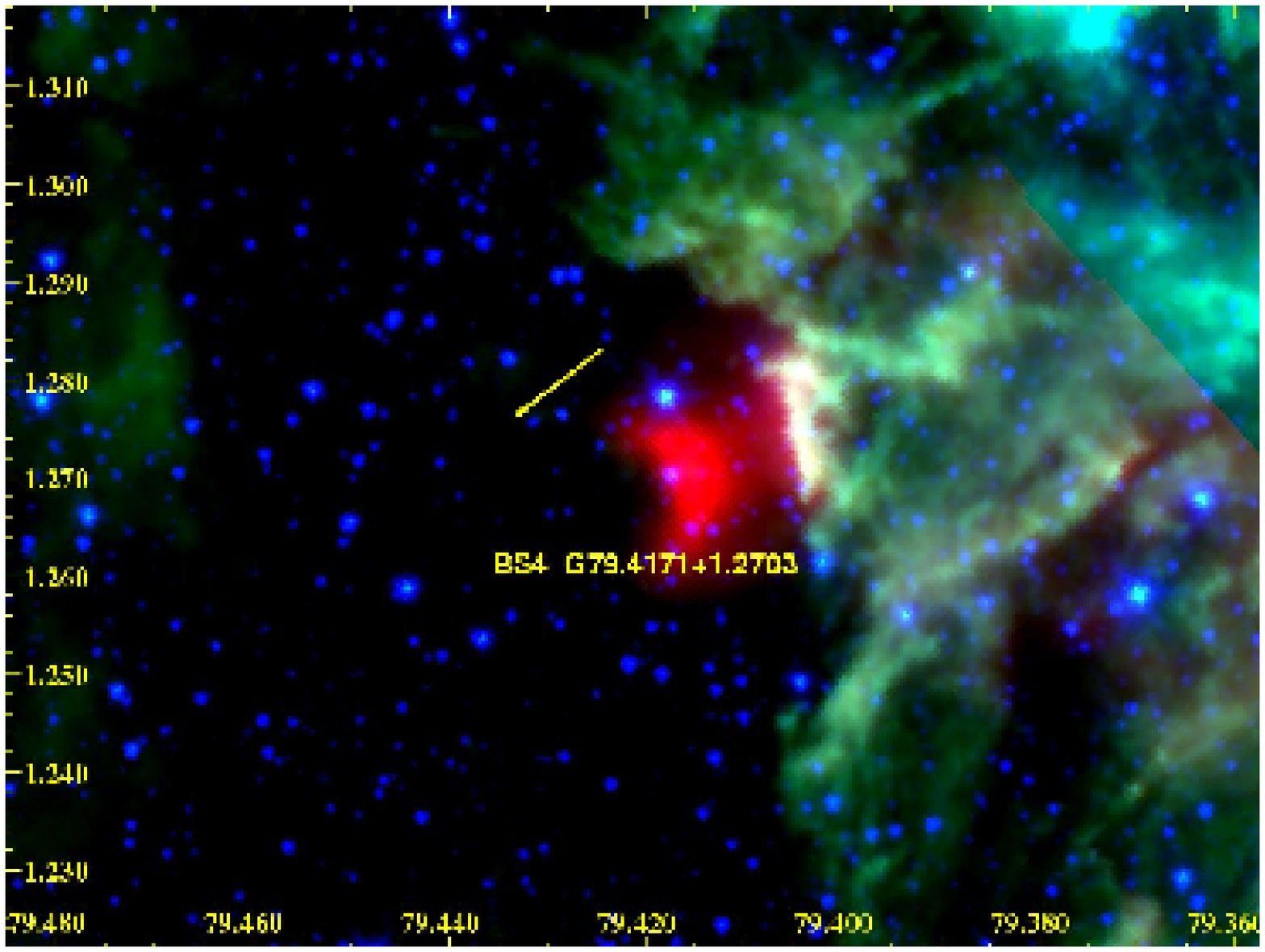}
\caption{Three-color image of Bowshock Candidate 4, as in Figure~\ref{overview}.
\label{BS4} }
\end{figure}

\begin{figure}
\plotone{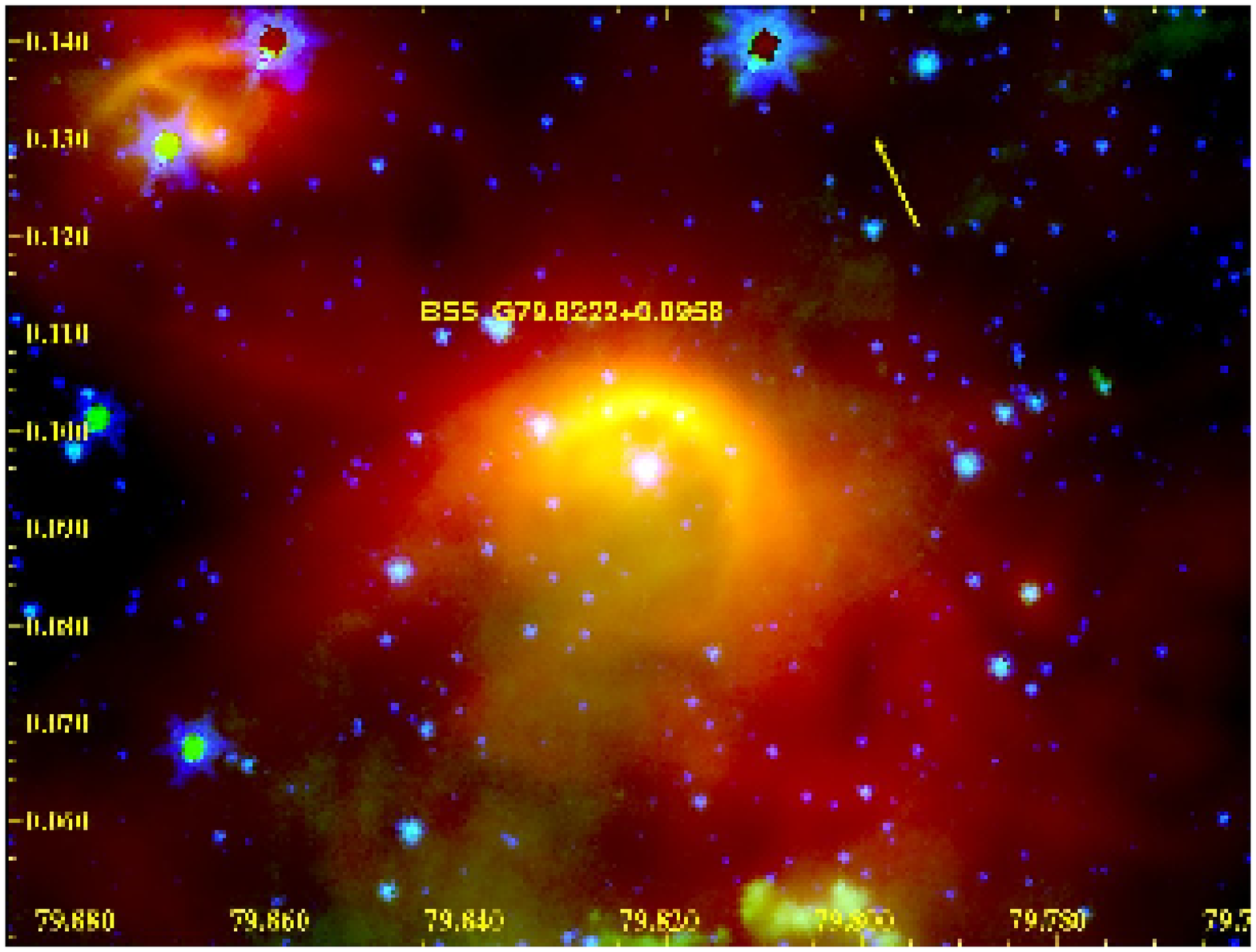}
\caption{Three-color image of Bowshock Candidate 5, as in Figure~\ref{overview}.
\label{BS5} }
\end{figure}

\begin{figure}
\plotone{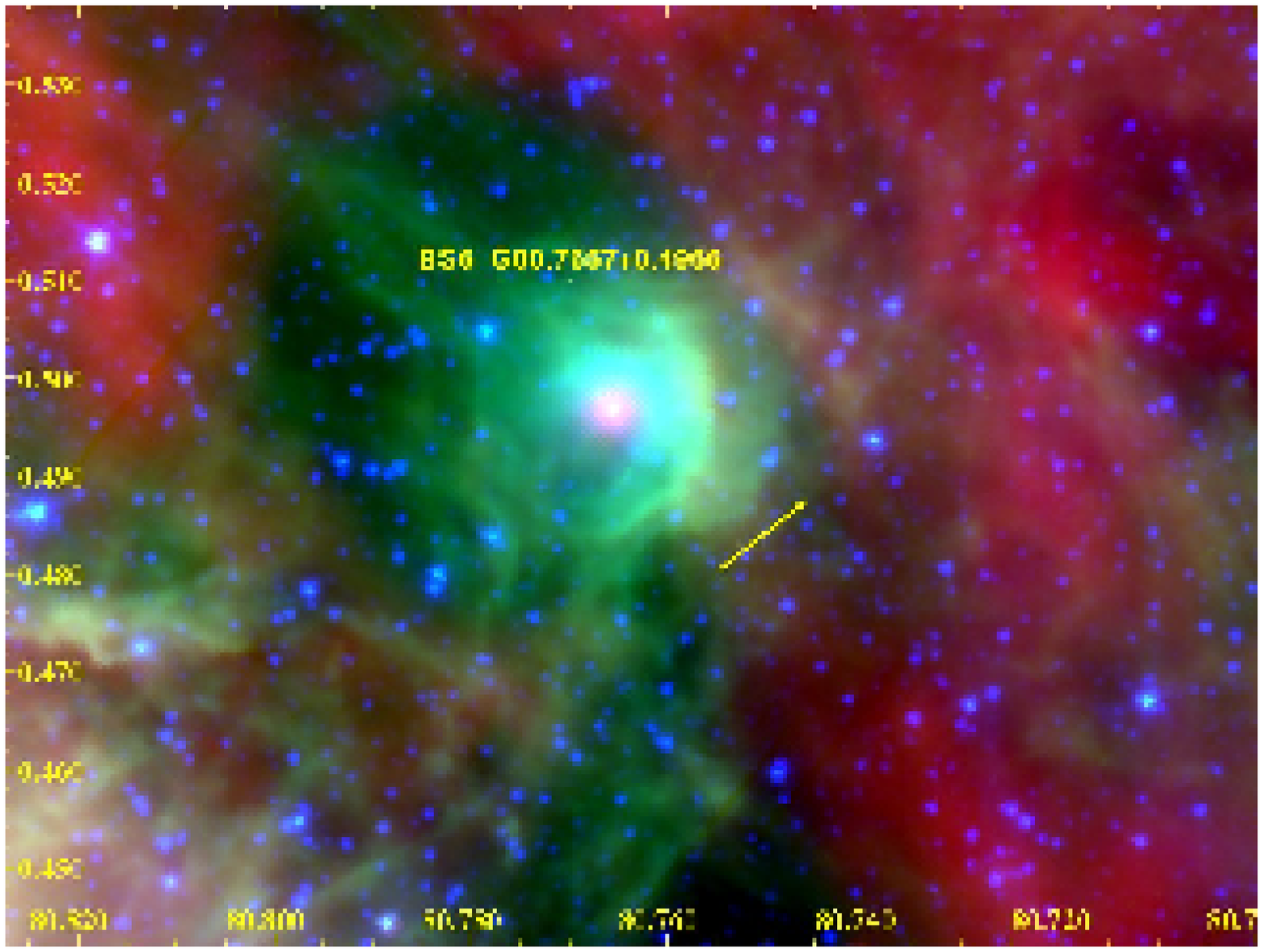}
\caption{Three-color image of Bowshock Candidate 6, as in Figure~\ref{overview}.
\label{BS6} }
\end{figure}

\begin{figure}
\plotone{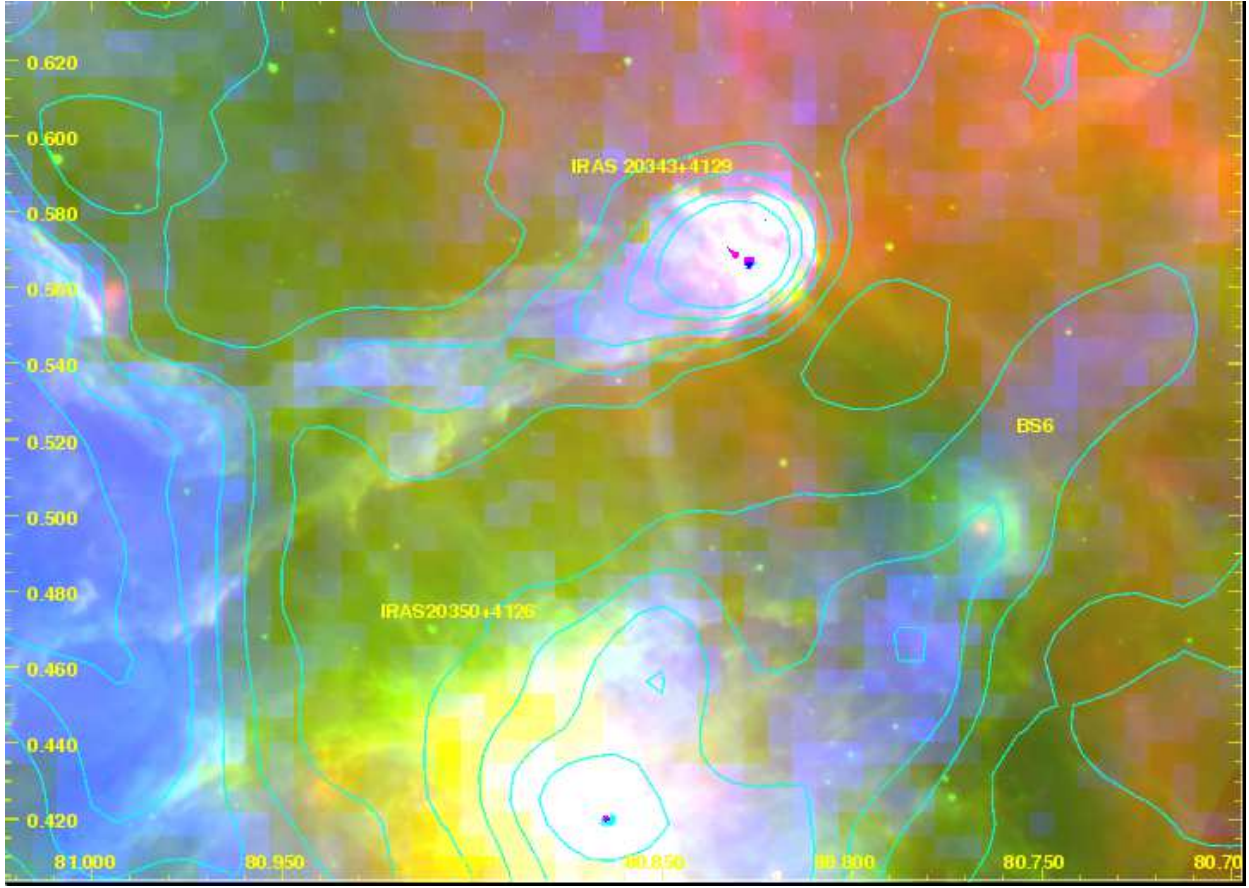}
\caption{Three-color image of Bowshock Candidate 6, with [8.0] in green, [24] in red,
and the $^{12}$CO (1-0) map between LSR velocities -7 -- -4 and 9 -- 14 \kms\
in blue and contours.  Star 6 and Bowshock Candidate 6 lie adjacent to a gasseous pillar
containing several massive young protostellar objects at its head (IRAS 20343+4129).  
\label{pillars} }
\end{figure}

HD195229 is also known as a high-proper-motion star with
$\mu_{RA}=0.14\pm0.55$ mas~yr$^{-1}$, and $\mu_{Dec}=-2.89\pm0.47$
mas~yr$^{-1}$ \citep{perryman}. Its {\it Hipparcos} proper motion
corresponds to a velocity of 28$\pm$8 \kms\ in the plane of the sky at
2000 pc. Combining this with the star's radial velocity gives a space
velocity of 32$\pm$8 \kms, placing it above the nominal 30 \kms\
threshold for runaway status.  The yellow bar in Figure~\ref{BS3}
shows the implied sky motion (from left to right) over 10,000 years
and has position angle uncertainty of about 15 degrees. The
orientation of the candidate bowshock implies a velocity vector toward
the lower left of Figure~\ref{BS3} while the proper-motion data imply
a velocity that is toward the lower right.
 
Figure~\ref{BSsed} shows that Bowshock 3 is detected with certainty
only at [24] and longer wavelengths, with a marginal detection at
[3.6] and [4.5].  The stellar SED is well fit by an early B giant
($T_{eff}$=25,000~K, R=10 $R_\odot$), consistent with its spectral
type, at a distance of 1.6 kpc with an extinction $A_V=1.0$ mag.
Curiously, HD195229 has a much lower extinction than the rest of the
stars in Cyg~OB2 (4--6 mag), requiring either that this star is seen
through a local minimum in the obscuring dust, or that HD~195229 is
actually on the near side of a heavy veil of extinction that enshrouds
most of Cygnus~X. 

The lack of similarity between the [24] appearance and the theoretical
shape in Figure~\ref{sim}, coupled with the misalignment between the
proper motion and putative bowshock morphology, casts doubt on a
bowshock interpretation.  We categorize the nature of this nebula as
ambiguous.


\subsubsection{Bowshock Candidate 4 and Star 4 (G79.4171+1.2703)}

Bowshock candidate 4 resembles Bowshock 1, having no detectable
emission at the IRAC bandpasses and a lima-bean morphology at
[24]. Note the edge-brightened cloud to the right of the bowshock,
suggesting illumination by the (presumed) early-B star powering the
nebula.  The central star is the faintest in our sample at optical
and infrared wavelengths and Figure~\ref{ccd} suggests a very large
extinction of $A_V>12$.  Its K-band magnitude is similar to the
central star of Bowshock 2, so it is probable that this is also an
early B star, if located at a similar distance.  There is no
literature identification for this star, so we give it a designation
according to its Galactic coordinate, G79.4171+1.2703.

We note that the putative bowshock nebula points away from Cyg~OB2 and in the
direction of the \HII\ region IRAS20264+4042, located toward the upper
right from Star 4 in Figure~\ref{overview}.  This star forming
region is also identified as the radio source DR7 \citep{dr} and the
host of the ultra-compact \HII\ region G79.320+1.313
\citep{kurtz94}. While it is possible that outflows from this SF
region are responsible for the high relative velocities leading to the
production of the putative bowshock (as shown for bowshocks in M17 and
RCW~49 by \citet{povich}), a careful search for other similarly
oriented bowshocks around IRAS20264+4042 revealed none.

 \citet{odenwald} report a $^{12}$CO detection for IRAS20264+4042 at
$V_{LSR}=-42$ \kms, corresponding to a kinematic distance of 6--7 kpc
inferred from the \citet{clemens} Galactic rotation curve.  However,
\citet{dutrabica01} and \citet{leduigou02} suggest distances of 1.1
and 1.6 kpc respectively, based on near infrared photometry.  At such
a distance, the K=11.3 magnitude of the central star, coupled with an
implied extinction of $A_K\sim$1.5 mag would correspond to an early-
to mid-B star.  At the the larger (but, we contend, less probable)
distance of 6.5 kpc, the apparent magnitude would be consistent with
an O9V or similar.  Given the abundance of early-type stars at the
nearer distance of Cyg~OB2, we prefer the near distance and the
corresponding early-to-mid B spectral type.
   
Figure~\ref{BSsed} shows that Bowshock Candidate 4 is detected only at
[24] and [70].  Although the stellar spectral type is not known with
certainty, the stellar SED is well fit by an $\sim$B3 dwarf
($T_{eff}$=20,000~K, R=4 $R_\odot$) at a distance of 1.6 kpc with a
very high extinction of $A_V=16$ mag.  The stellar photometry exhibits
an excess at [8.0], suggesting the presence of hot dust or possibly
PAHs associated with the star.  Lack of similarity to the theoretical
bowshock shape leads us to classify this object as ambiguous.

\subsubsection{Bowshock Candidate 5 and Star 5 (A10)}

Figure~\ref{BS5} shows Bowshock Candidate 5 to be a stunning
combination of emission at the IRAC and MIPS bandpasses.  The central
star is A10 in the notation of \citet{comeron02}. Our new spectrum of
A10, shown in Figure~\ref{specA10}, is consistent with a late O star,
although the uncertainty is several spectral subclasses.  Its
heliocentric radial velocity is 10$\pm$10 \kms, consistent with other
Cyg~OB2 members.  Its JHK colors and magnitudes are similar to the
central star of Bowshock Candidate 1, suggesting a similar reddening and late O
spectral type if at the distance of Cyg~OB2.  The angular distance
between the star and bowshock apsis is $\sim$25\arcsec\ or 0.19 pc
projected separation.

Figure~\ref{BSsed} shows that the 2MASS data are well fit by an O9V
($T_{eff}$=31,000~K, R=8 $R_\odot$) at a distance of 1.6 kpc with an
extinction of $A_V=7.5$ mag.  The nebula dominates the SED at the
IRAC and MIPS bandpasses.

Figure~\ref{sim} shows that the [8.0] and [24] appearance 
closely resembles the theoretical bowshock shape.  The orientation 
is consistent with either motion {\it toward} Cyg~OB2 or 
interaction with a flow of material emanating from Cyg~OB2.  
We classify this object as a probable bowshock.  


\subsubsection{Bowshock Candidate 6 and Star 6 (G80.7657+0.4966)}

Bowshock candidate 6 exhibits strong emission at both IRAC and MIPS
bandpasses in Figure~\ref{BS6}.  It lies within a complex of diffuse
emission that may constitute a small \HII\ region.  The central star
is extremely red, and Figure~\ref{ccd} suggests an extinction
approaching $A_V=18$.  Accordingly, there is no literature
identification or spectral type available, so we designate it by its
Galactic coordinates, G80.7657+0.4963. 

Though the nebular component of object 6 is strong at the IRAC and MIPS
 bandpasses, the central star is the faintest in the sample at IRAC
 bandpasses.  While there is no spectral information on Star 6, if we
 assume it is at 1.6 kpc the 2MASS SED is well fit by a mid-B dwarf
 ($T_{eff}$=15,000~K, R=3 $R_\odot$) with a very high extinction of
 $A_V=$19 mag.

Star 6 is one of only three detected at [24], suggesting the presence
of a dusty shell or disk associated with the star, since we would not expect to
detect the stellar photospheres at [24] or [70]. Figure~\ref{BSsed}
shows that the flux from the star at [24] is greater than the nebular
flux and continues to rise beyond 24 microns.  These characteristics
suggest that Star 6 is a pre-main-sequence object illuminating a rim of material
that points toward Cyg~OB2, the likely source of a high-velocity wind.

Figure~\ref{pillars} shows the [8.0] image (green), [24] image (red),
and $^{12}$CO (1-0) molecular line map (blue and contours) integrated
between LSR velocities -7 -- -3 \kms\ and 9 -- 14 \kms \citep{brunt}.
This figure shows that Bowshock 6 is located along a tendril of CO (at
velocities -7 -- -3 \kms) pointing toward Cyg~OB2.  It lies adjacent
to a prominant gasseous ``pillar'' seen at [8.0] and in CO.  The star
forming region IRAS~20343+4129, which contains several
pre-main-sequence B stars \citep{campbell2008}, constitues the head of
this pillar and is coincident with a concentration of CO at velocities
9 -- 14 \kms.  This pillar also points toward Cyg OB2, consistent with
it being eroded by radiant and mechanical luminosity of the massive
stars therein.  The similarity between the contents (young early B
stars) and morphology (pillar/CO filament) of the Star 6 region and
IRAS~20343+4129 raises the intriguing possibility that these two
represent different phases, and perhaps different viewing angles of
the same phenomenon: massive star formation triggered in the heads of
gas columns being eroded from the outside.  The lack of a strong CO
concentration and visible pillar of photo-dissociated material near
Star 6, coupled with the presence of the bowshock pointing toward
Cyg~OB2 suggests that the region containing Star 6 may be at a more
evolved stage compared to IRAS~20343+4129. It may also be that the
stars within IRAS~20343+4129 are more massive and, thereby, generate a
more luminous \HII\ region.  We classify the central star of Bowshock
Candidate 6 as a probable young stellar object.
  
\subsubsection{Bowshock Candidate 7 and Star 7 (A37)}

\begin{figure}
\plotone{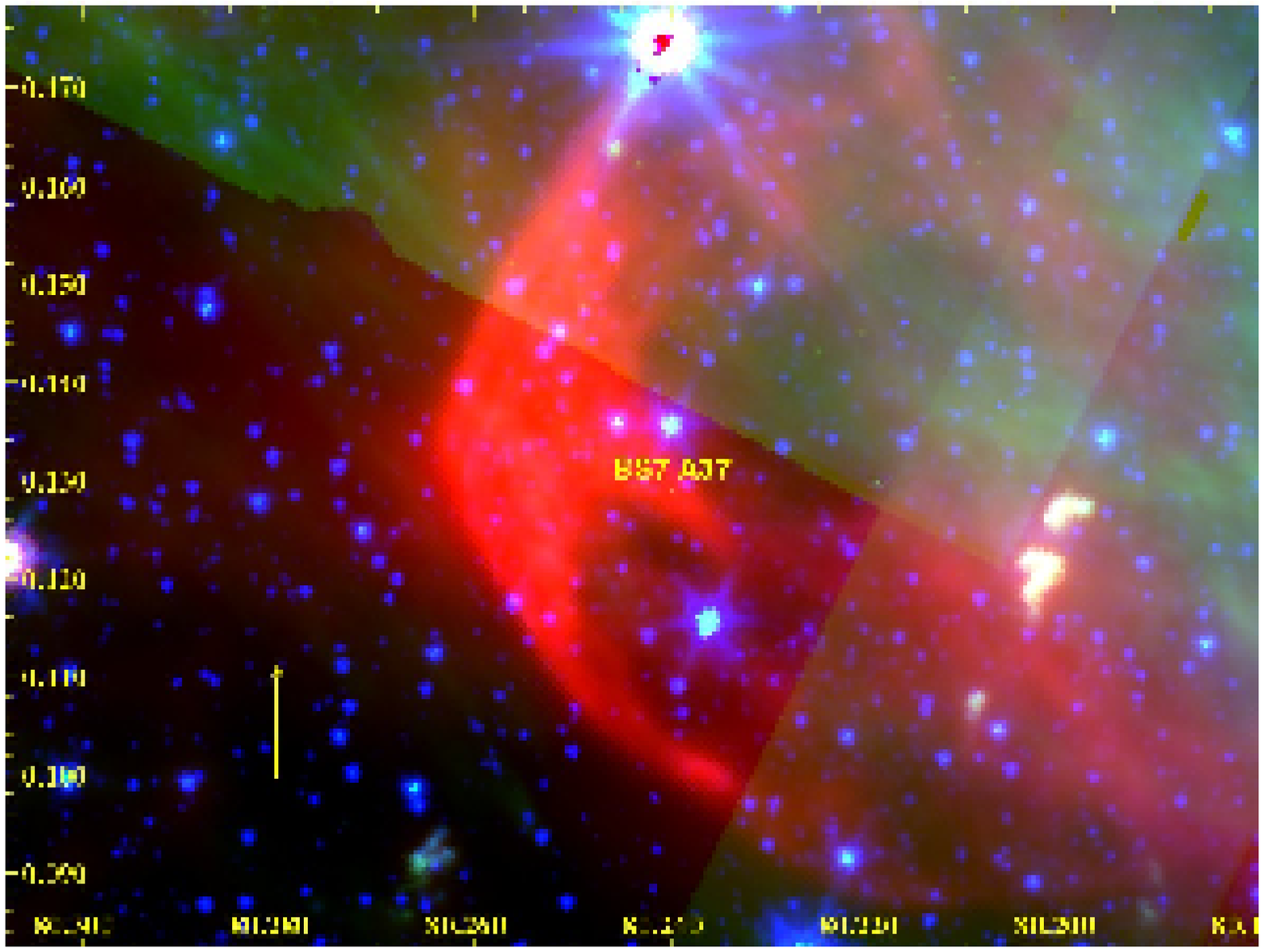}
\caption{Three-color image of Bowshock Candidate 7, as in Figure~\ref{overview}.
\label{BS7} }
\end{figure}

\begin{figure}
\plotone{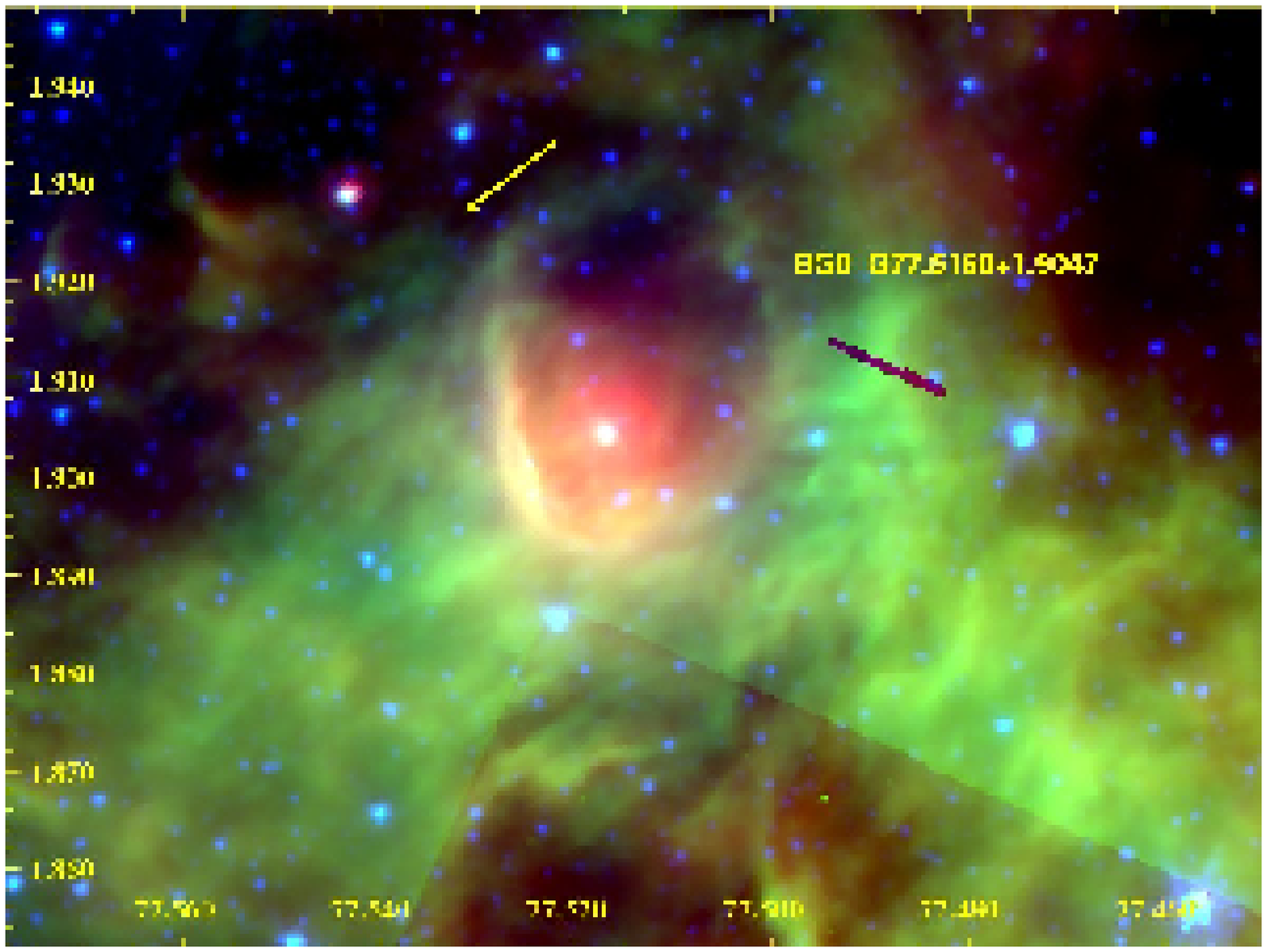}
\caption{Three-color image of Bowshock Candidate 8, as in Figure~\ref{overview}.
\label{BS8} }
\end{figure}

\begin{figure}
\plotone{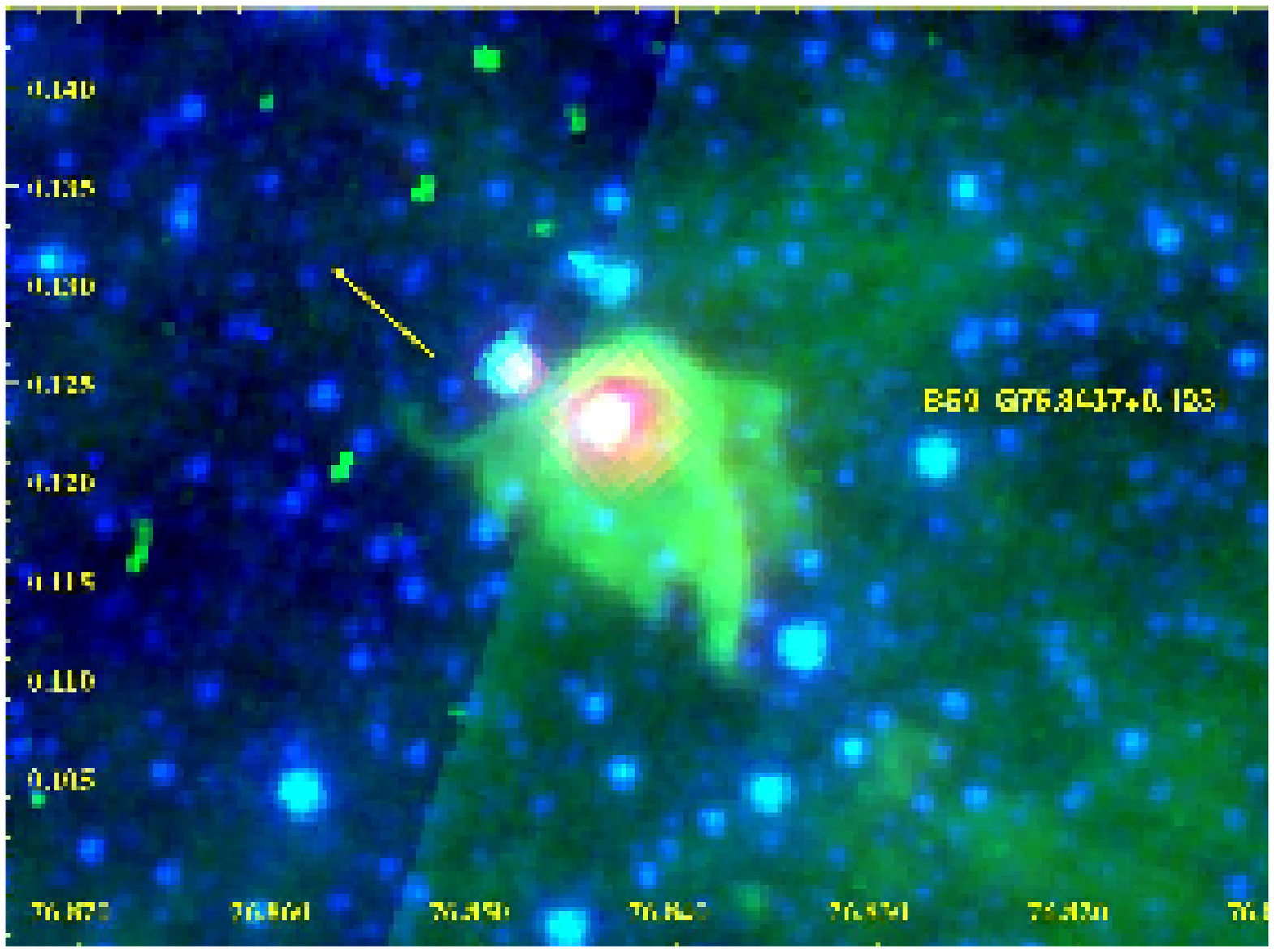}
\caption{Three-color image of Bowshock Candidate 9, as in Figure~\ref{overview}.
\label{BS9} }
\end{figure}

\begin{figure}
\plotone{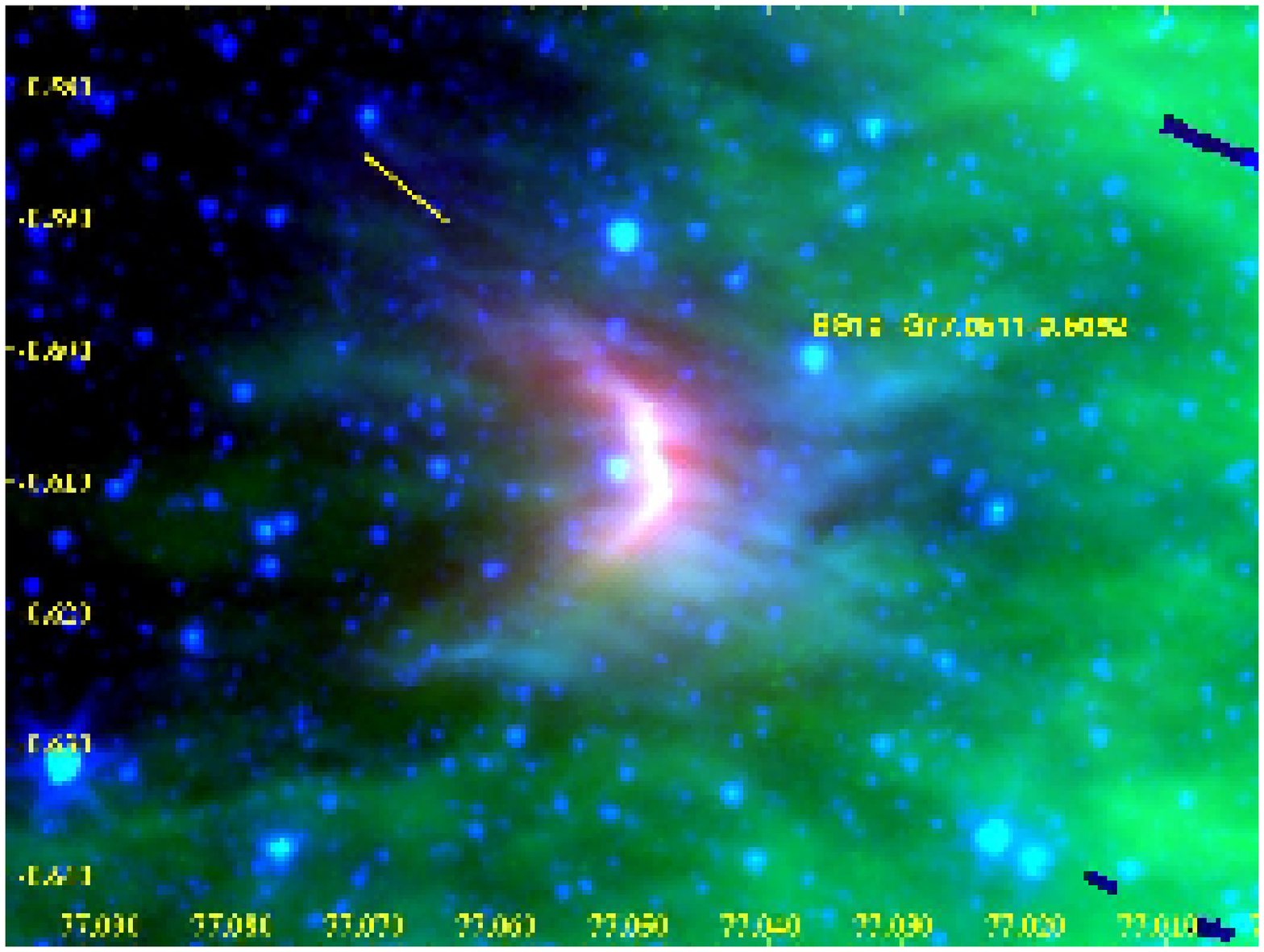}
\caption{Three-color image of Bowshock Candidate 10, as in Figure~\ref{overview}.
\label{BS10} }
\end{figure}

\begin{figure}
\plotone{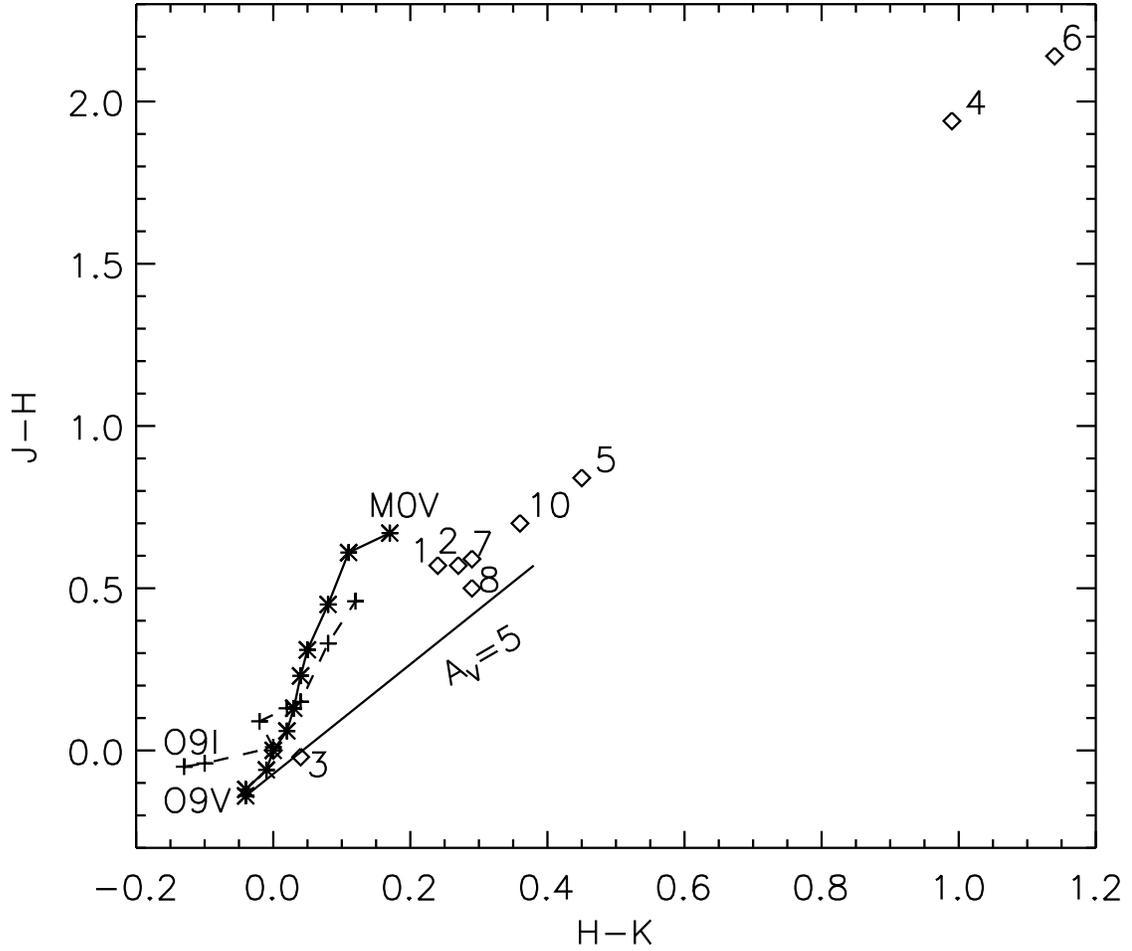}
\caption{A JHK color-color diagram showing the sample stars ({\it diamonds}) in
relation to the fiducial main sequence and supergiant sequence
({\it solid line} and {\it dashed line}). The reddening vector shows the equivalent of 
5 magnitudes of visual extinction based on the \citet{cardelli89} reddening curve.  
\label{ccd} }
\end{figure}

Bowshock candidate 7 has the largest angular extent of any of our targets,
covering nearly 5\arcmin\ in Figure~\ref{BS7}.  It is detected only at
[24] and [70].  The central star appears to be A37 in the notation of
\citet{comeron02}.  \citet{hanson03} determines a spectral type of
O5V((f)), making it the most energetic and massive in our sample.  
The presence of a bowshock associated with the this star is
noted by \citet{bomans08} on the basis of images from the {\it
Midcourse Space Experiment} mission.  The angular distance between the
star and bowshock apsis is the largest in our sample, $\sim$70\arcsec,
or 0.53 pc projected separation.

The SED of Star 7 in Figure~\ref{BSsed} is consistent with an
O5V star ($T_{eff}$=40,000~K, R=11 $R_\odot$) and $A_V=5$ mag if
placed at a distance of 2.1 kpc, slightly larger than the nominal
distance to Cyg OB2, although the line-of-sight depth along the
Cygnus-X complex is likely to be $\sim$few hundred pc \citep{odenwald}.

The excellent agreement between the [24] morphology and the
theoretical shape leads us to classify this as a probable bowshock.
The orientation of the bowshock indicates a motion toward the left in
Figure~\ref{BS7}, roughly tangent to the vector toward Cyg~OB2.
This suggests that the origin of A37 lies somewhere outside of Cyg~OB2.

\subsubsection{Bowshock Candidate 8 and Star 8 (G77.5168+1.9047)}

Bowshock candidate 8 is visible at both IRAC and [70] bandpasses.
Figure~\ref{BS8} shows that this object lacks the classical arc
morphology and symmetry of the other bowshocks, warranting an
ambiguous classification.  It resembles a partial shell or bubble.
 The WIRO spectrum of this star obtained
2009 September 16 is consistent with a B2V -- B3V having a
heliocentric radial velocity of 2$\pm$4 km~s$^{-1}$, similar to
other Cyg OB2 members.  The apparent central star, designated
G77.5168+1.9047, is relatively bright at J=10.5, and it has 2MASS
magnitudes and reddening consistent with other bowshock stars and
early-type Cyg~OB2 members. 

Figure~\ref{BSsed} shows that the near-IR SED of Star 8 is well fit by
a B3V ($T_{eff}$=20,000~K, R=4 $R_\odot$) with $A_V=4.5$ at a distance
of 1.6 kpc.  The IRAC and MIPS bands are dominated by
extra-photospheric emission, consistent with hot dust near the star.
Star 8 is one of only three stars (Star 6 and Star 9) detected at
[24], suggesting the presence of dust in a circumstellar disk or
envelope.  The morphology of the nebula is does not bear a strong
resemblance to the theoretical bowshock shape.  As such, we designate
Star 8 as a possible young stellar object (YSO).

\subsubsection{Bowshock Candidate 9 and Star 9 (G76.8437+0.1231)}

Bowshock candidate 9 is a compact nebula with detected emission only
at [5.8] and longward.  The central star is clearly seen as a point
source in Figure~\ref{BS9}. 
The stellar SED of Star 9 is difficult to ascertain, since there is no
2MASS counterpart to the point-like source detected at the mid-IR IRAC
bands.  The mid-IR flux rise steeply with wavelength, consistent with
a heavily enshrouded source.  The [4.5] flux lies below the
expectations of a simple interpolation between the [3.6] and [5.8]
bands.  This has become a classic signature of material dominated by
emission from large molecules (e.g, PAHs; \citet{draineli07}), which
are largely absent in the [4.5] band.  The nebular morphology does not
bear a strong resemblance to the classical bowshock shape.  These
characteristics are consistent with a young (class I?) protostar
\citep{kenyon93}, and we designate this object as a probable YSO.

\subsubsection{Bowshock 10 and Star 10 (GG77.0511-0.6092)}

Bowshock 10, pictured in Figure~\ref{BS10} is a small, narrow filament
with strong emission in the IRAC bands, and a low ratio of far-IR to
mid-IR flux in the final panel of Figure~\ref{BSsed}.  The stellar
source is identified only by its Galactic coordinates,
G77.0511-0.6092.  Because it has similar 2MASS colors and magnitudes
as many of the other bowshock stars, we consider it to be a probable
late O or early B star at the distance of Cyg~OB2.

The near-IR SED of Star 10 is consistent with an early B star
($T_{eff}$=25,000~K, R=5 $R_\odot$) with $A_V=6$ at a distance of 1.6
kpc.  The IRAC points lie in excess of the photosphere, consistent
with hot dust and emission from PAHs.   The morphology of the [8.0] nebula
appears irregular, while the [24] nebula is similar to the classical
bowshock morphology.  We classify this object as ambigous.

\subsubsection{BD+43\degr3654}

\citet{comeron07} present proper motions and {\it MSX} infrared images
of the O4If star BD+43\degr3654, originally identified as a probable
runaway by \citet{vanb88} on the basis of a probable bowshock seen
with IRAS.\footnote{BD+43\degr3654 lies outside the area of the {\it
Spitzer} Legacy Survey of the Cygnus-X Complex, so we do not discuss
this object in detail or present figures.}  They compute a distance of
1450 pc and propose that BD+43\degr3654 was ejected from Cyg~OB2 1.6
Myr ago, not long after its birth.  \citet{bomans08} propose that
BD+43\degr3654 is a blue straggler, formed from the merger of two
stars and ejected from Cyg~OB2 during a binary-binary encounter. They
invoke the high proper motion pulsars B2020+28 and B2021+51 as the
descendants of the other two stellar participants in this 4-body
interaction.

We cross-correlated the WIYN spectrum of BD +43\degr3654 with an O5I
model atmosphere, which yielded a heliocentric radial velocity of
$V_{LSR}$=$-$66.2$\pm$9.4 \kms.  This is considerably more negative
than the $V_{LSR}$=$-$10 \kms\ mean velocity of O stars in Cyg~OB2
\citep{kiminki07}, and therefore BD +43\degr3654 would qualify as a
runaway on the basis of its radial velocity alone.  Its motion in the
plane of the sky is 39.8 $\pm$ 9.8 \kms\ \citep{comeron07} at the
assumed distance. Combining this with our radial velocity measurement,
we calculate the heliocentric space velocity of BD+43\degr3654 to be
77 $\pm$ 10 \kms.

\subsection{Summary of Classifications and Statistics of Runaways in Cygnus~X}

The bowshock candidates presented here comprise a mixed
bag of phenomena.  Objects 2, 5, and 7 (and BD +43\degr3654) are most
likely to be genuine bowshocks powered by runaway stars on the basis
of having early spectral types and nebular morphologies most similar
to the classical bowshock appearance.  However, none of stars 2, 5 or
7 have remarkable radial velocities.  Objects 1, 3 (the high
proper-motion star HD195299), 4, and 10 are ambiguous, having nebular
morphologies less similar to the classical bowshock shape, but,
nevertheless, appear to be early-type stars.  Objects 6, 8, and 9 have
central stars with excess IR emission, suggesting circumstellar
material.  Stars 6 and 9 also have high extinction and/or SEDs that
rise toward longer wavelengths, consistent their being young stellar
objects hosting disks or circumstellar envelopes.  The nebula
associated with Star 6 shows evidence for being the head of a gaseous
pillar being ablated by radiation from the direction of Cyg~OB2.
Hence, objects 6, 8, and 9 are the least likely to be high velocity
runaway stars.

In a magnitude-selected sample of 195 bright O stars, \citet{gies86}
found that 16 (8\%) could be classified as runaways either on the
basis of radial velocities (6; 3\%), proper motions (5; 2.5\%), or
extreme distance from the Galactic Plane (5; 2.5\%).  By contrast the
radial velocity survey of 146 Cyg~OB2 stars by \citet{kiminki07} did
not find any runaways.  The {$SST$} Cygnus~X Legacy Survey covers a
much larger region ($\sim23$ square degrees) than the sample of core
Cyg~OB2 members studied by \cite{kiminki07} ($\sim0.35$ square
degrees), which was selected from the optical imaging survey of
\citet{mt91}.  The ratio of areas in these two surveys is roughly 65,
so one might conclude that the number of early type stars in Cygnus~X
is as many as 146 $\times$65 = 9500.  However, the earliest and most
massive stars appear to be contained within the core Cyg~OB2 region
covered by \cite{mt91}, so that this is a gross overestimate.  We very
conservatively adopt that the $SST$ Cygnus~X Legacy Survey survey
includes $\sim$600 OB stars earlier than B2, or about four times the
number of OB stars surveyed by \citet{kiminki07}.  In this region we
find only three candidates for runaways---four if BD~+43\degr3654 is
included.  At face value, this is 3/600 = 0.5\%.  Of course, high
space velocities are a necessary but not a sufficient condition for
for the formation of bowshocks.  \citet{vanb95} found that 58 out of
188 high velocity stars evince bowshocks at IRAS sensitivity levels,
and of these, 25 have well resolved structure.  We expect that {\it
SST} would detect even fainter and smaller bowshocks.  If we assume
that 1/4 of runaway stars produce bowshocks visible at {\it SST}
sensitivity levels and that 8\% of all OB stars are runaways, we would
expect to find on the order of 12 bowshocks in the Cygnus~X region.
Why do we detect so few?

We identify only three probable bowshock runaways, and perhaps
only one of these (Bowshock 2) has an orientation suggesting an origin
near Cyg~OB2.  It is certainly possible that most of the runaways,
like BD~+43\degr3654, could fall outside the survey area if they have
high space velocities and wereprobable ejected from the Association very early
in its formation.  Stars with a tangential velocity of 30 \kms\ would
travel beyond the boundaries of the {\it SST} Cygnus~X Legacy Survey
($\sim$26 pc diameter) in about 0.5 Myr---a small fraction of the 2--4
Myr age of Cyg~OB2.  Dynamical ejection scenarios wherein N-body
interactions in close multiple systems expel stars would predict
this---that most of the runaways are produced early in the
formation of a young massive star cluster and have already traveled 
beyond the immediate cluster vicinity.  Supernova ejection
scenarios, on the other hand, predict that the start of the ejection
process is delayed several Myr (until the explosion of the most
massive stars), and then it continues at a constant or increasing rate
as stars from lower-mass OB binaries complete their evolution.
Although the arguments are mostly qualitative, our findings would seem
to be more consistent with the dynamical ejection scenario.

\section{Bowshocks and Laboratories}

\citet{baranov} first articulated the theoretical basis for
momentum-driven bowshocks in the context of the solar wind termination
shock.  The theory of stellar wind bowshocks has since been further
refined in papers by \citet{vanb88}, \citet{vanb90}, \citet{wilkin96},
\citet{wilkin00} and simulated numerically by \citet{comeron98}.  A
bowshock forms at a radial distance $R_0$ where the momentum flux from
the stellar wind material of density $\rho_w$ and velocity $V_w$ is equal to the
momentum flux from the ambient medium of density $\rho_a$ moving at
$V_a$ relative to the star.  The relative velocity may be the result
either of a ``runaway'' star moving at high speed or a bulk flow of
the ISM, such as an outflow from a young star-forming region.  Hence,
 
\begin{equation} 
{1 \over 2} \rho_{w} V_w^2 = {1 \over 2} \rho_{a} V_a^2.
\end{equation}

\noindent The density of the stellar wind can be expressed as

\begin{equation}
\rho_w = {\dot{M_w} \over {4\pi R_0^2 V_w}} ~ ~ ,
\end{equation}

\noindent where $\dot{M_w}$ is the stellar mass loss rate.  It follows that the
 standoff distance of the bowshock is 

\begin{equation}
R_0 = \sqrt {{V_w \dot{M_w}} \over {4\pi \rho_a V_a^2}}  ~ ~ .
\end{equation} 

$R_0$ can be measured directly from the bowshock images, assuming that
all of the stars are located at the adopted distance of Cyg~OB2.
Technically, only the projected separation, $R_0~cos~i$, can be
measured, but in order to observe a bowshock morphology, the viewing
angle cannot be far from $i=0^\circ$ so that $cos~i$ is not far from
unity.  Adopting a mean ISM gas mass per H atom of
$\mu=2.3\times10^{-24}$~g means that $\rho_a$ can be expressed in
terms of the ambient number density: $\rho_a = \mu n_a$.  Stellar wind
speeds and mass loss rates have proved difficult to measure and may be
even more uncertain than previously believed \citep{puls08,
fullerton06}.  As an estimate of the mass loss rates and wind
velocities for stars in our sample, we adopt the values tabulated by
\citet{mokiem07} for early type stars.  Equation 3 can then be
rearranged to yield the relative motion btween the star and ambient
medium,

\begin{equation}
 V_a = \sqrt{{V_w \dot{M_w}} \over {4\pi R_0^2 \mu}} n_a^{-1/2} ~ .
\end{equation}

In practice, it is difficult to calculate the star's velocity, $V_a$,
relative to the surrounding medium because the quantities on the right
hand side are poorly known, especially the density of the ambient
medium, $n_a$.  By adopting published estimates for the stellar mass
loss rates and wind velocities, we can at least make an estimate of
the product $V_a ~ n_a^{1/2}$, similar to the approach of
\citet{povich}.  In keeping with the precedent established in
\citet{vanb88} and \citet{povich}, we write the mass loss rate in units
of $10^{-6}$ $M_\odot$ yr$^{-1}$ as $\dot M_{w,-6}$, the stellar wind
velocity in units of $10^{8}$ cm s$^{-1}$ as $V_{a,8}$ and the ambient
medium density in units of $10^{3}$ cm$^{-3}$ as $n_{a,3}$.  The
relative star-ISM velocity is then,

\begin{equation}
V_a = 1.5 \bigl({{R_0}\over{pc}} \bigr)^{-1} (V_{w,8} ~ \dot M_{w,-6} )^{1/2} n_{a,3}^{-1/2} ~~[km~s^{-1}] . 
\end{equation}

Given the uncertainties in stellar mass loss rates and wind velocities
for stars later than about B0, we compute and tabulate derived
parameters only for the five objects in our sample with the most
reliable spectroscopic types, objects 1, 3, 5, 7, and 11
(BD+43\degr3654).  Table~\ref{tab.params} tabulates the adopted stellar
wind velocities, mass loss rates (from the works of \citet{repolust04,
mokiem05, martins05, crowther06}, as summarized in \citet{mokiem07}),
standoff distances, $R_0$ $cos~i$, along with the product $V_a ~
n_{a,3}^{1/2}$.  The standoff distances of 0.11 -- 1.8 $cos~i$ pc are
consistent with those found for the bowshocks in \citet{povich}.  The
derived values for $V_a ~ n_{a,3}^{1/2}$ lie in the range 0.2 to as
much as 30 --- broadly consistent with the values from the
\citep{povich} stars.  For typical ISM densities of $n_{a,3}$=0.1 --
1.0, the implied stellar speeds are few$\times10$ \kms, consistent
with the expected speeds of runaway stars.

\subsection{A Novel Measure of Stellar Mass Loss Rates  }

Because stellar mass loss rates are probably the most uncertain parameter
in Equation 5 above, we invert the approach taken by \citet{povich}
and exploit the phenomenon to yield an estimate of the
mass loss rates for the central stars.  Inverting Equation~5 yields the
mass loss rate of the star as a function of the relative motion between the
str and ISM, the ambient density,
the stellar wind speed, and the standoff distance, 

\begin{equation}
 \dot M_{w,-6} = {{0.67 [R_0(pc)]^2 [V_a(km/s)]^2  n_{a,3}} \over  { V_{w,8}}} ~ . 
\end{equation}

\begin{figure}
\epsscale{0.7}\plotone{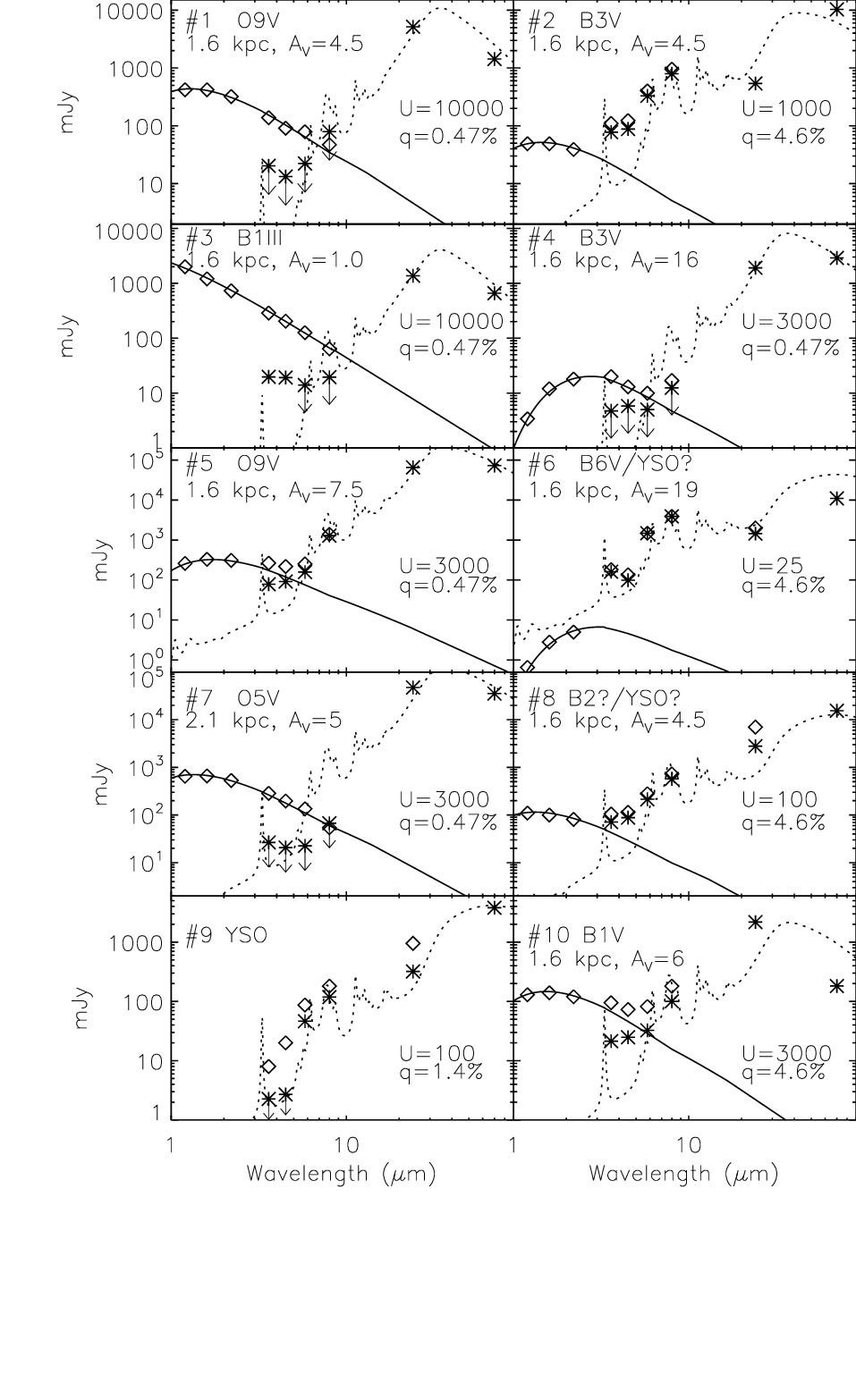}
\caption{Infrared spectral energy distributions of the sample nebulae based on the
photometry from Table~\ref{phot.tab}.     
\label{BSsed} }
\end{figure}

\begin{figure}
\plottwo{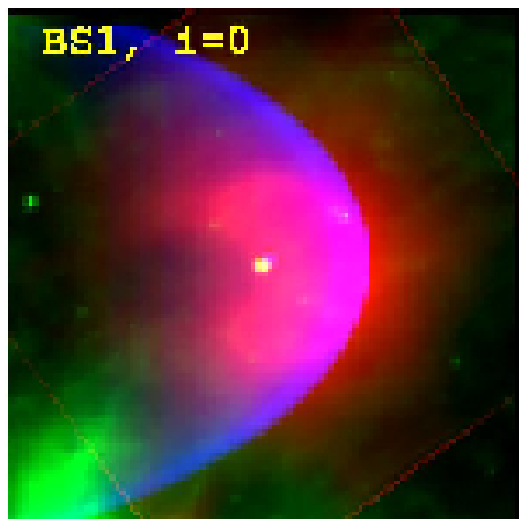}{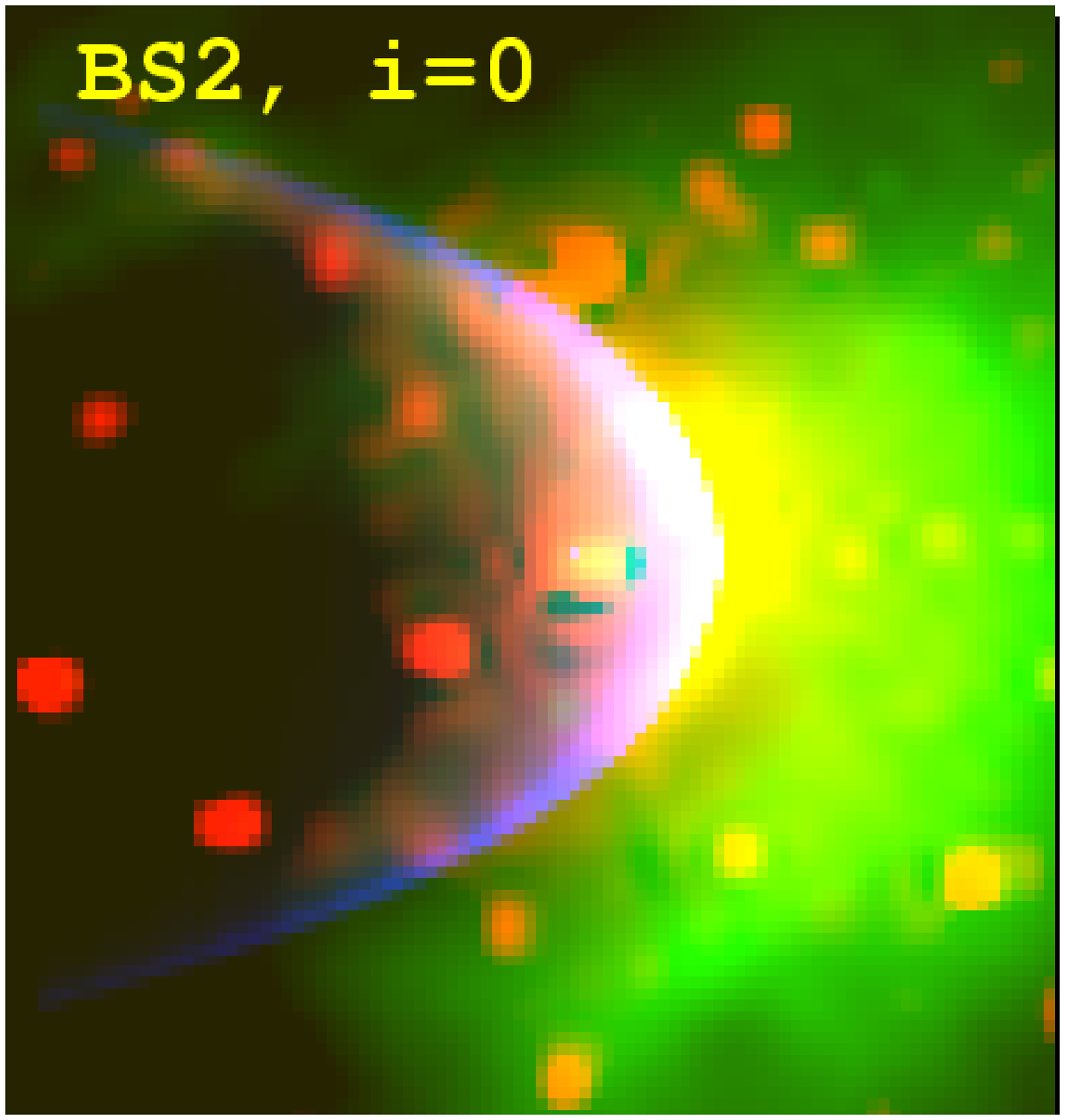}
\plottwo{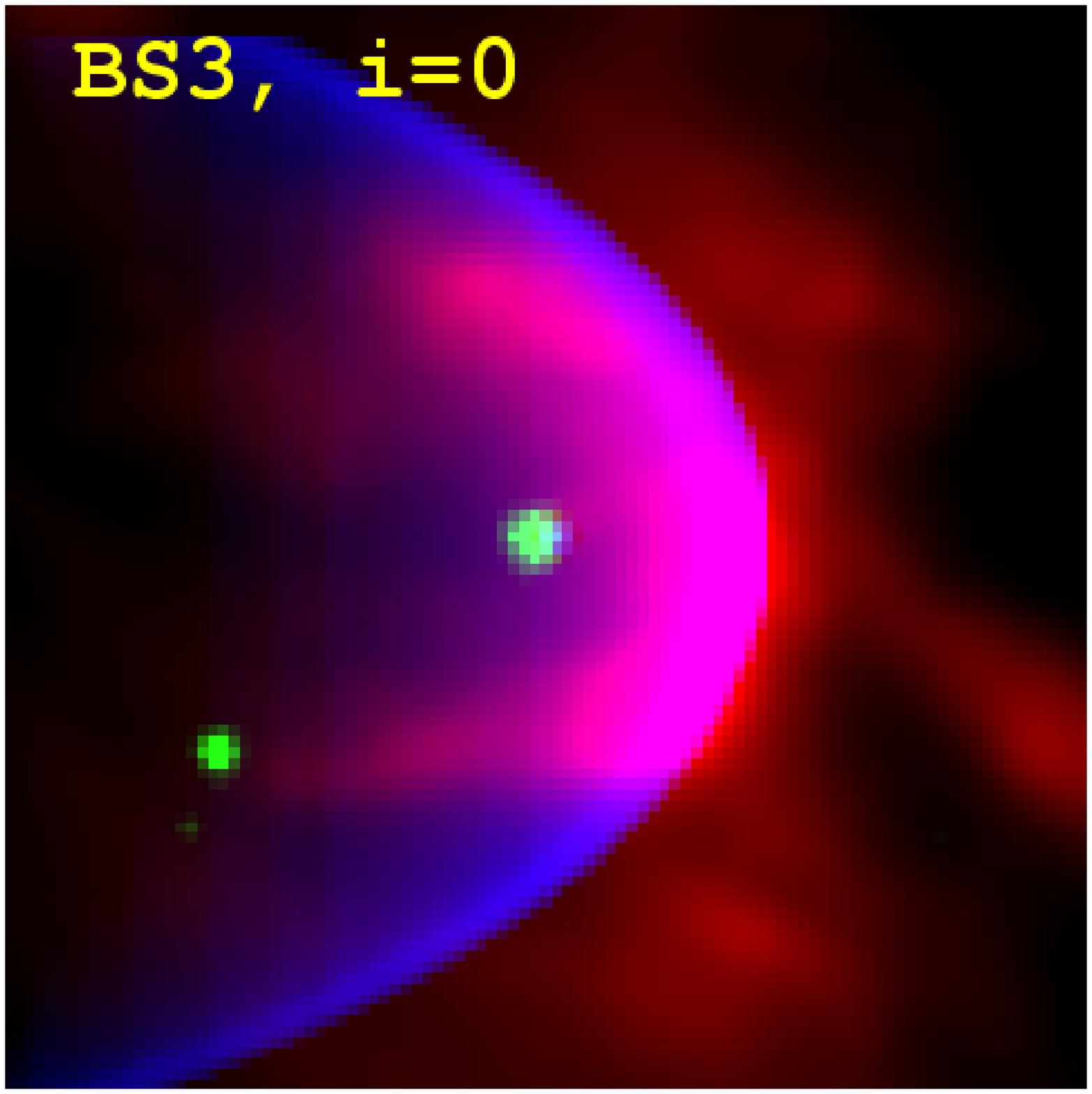}{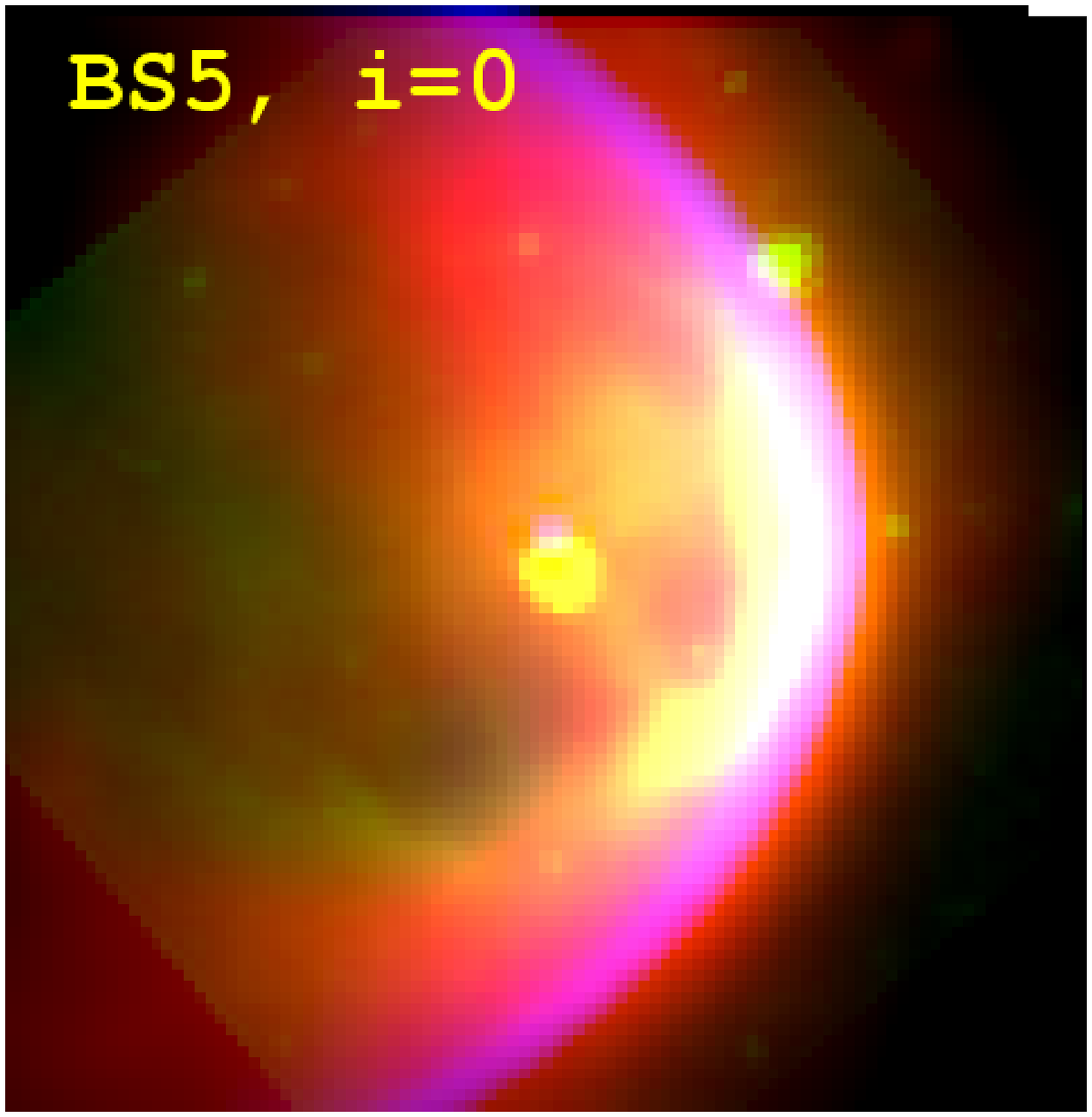}
\plottwo{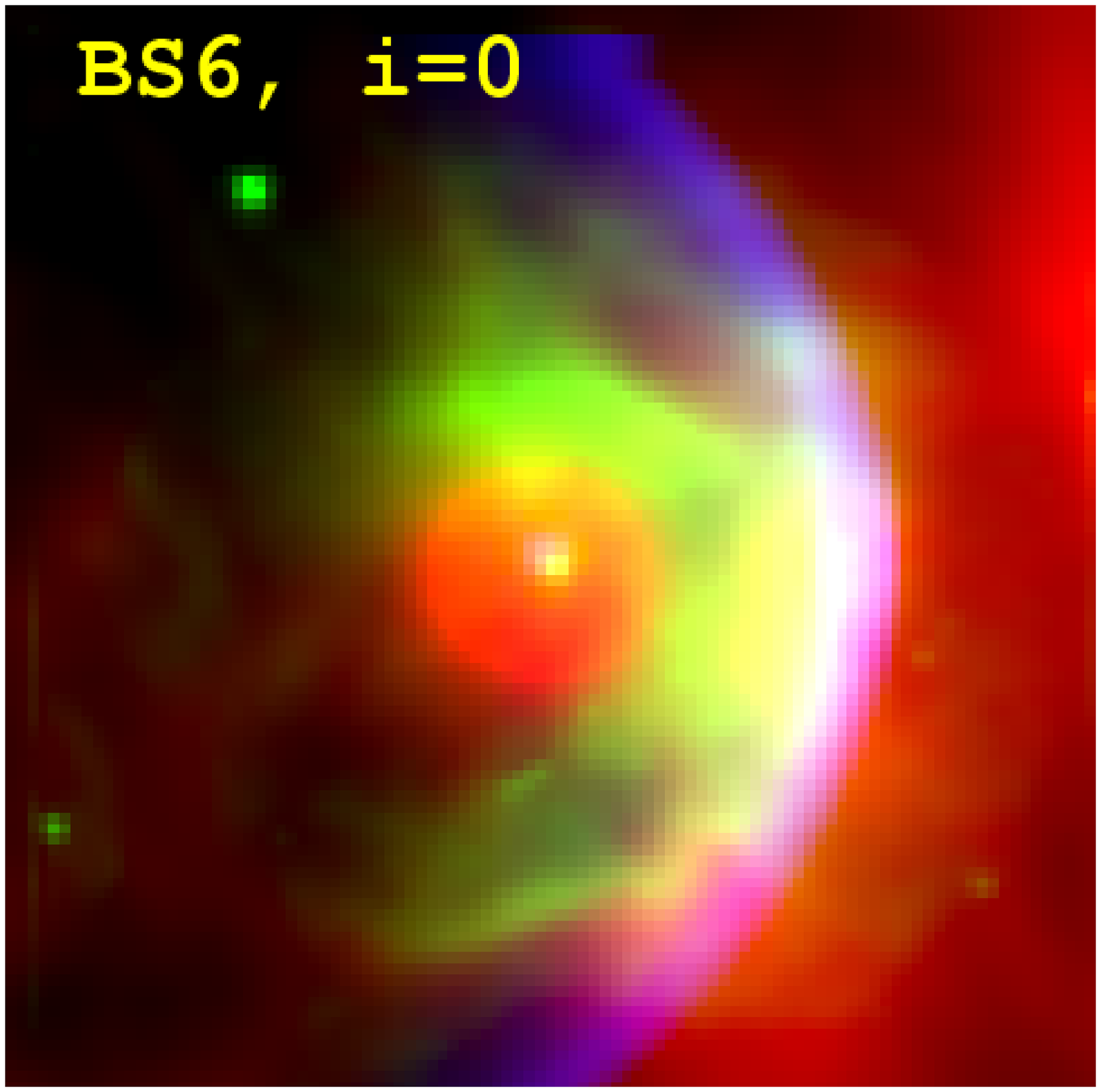}{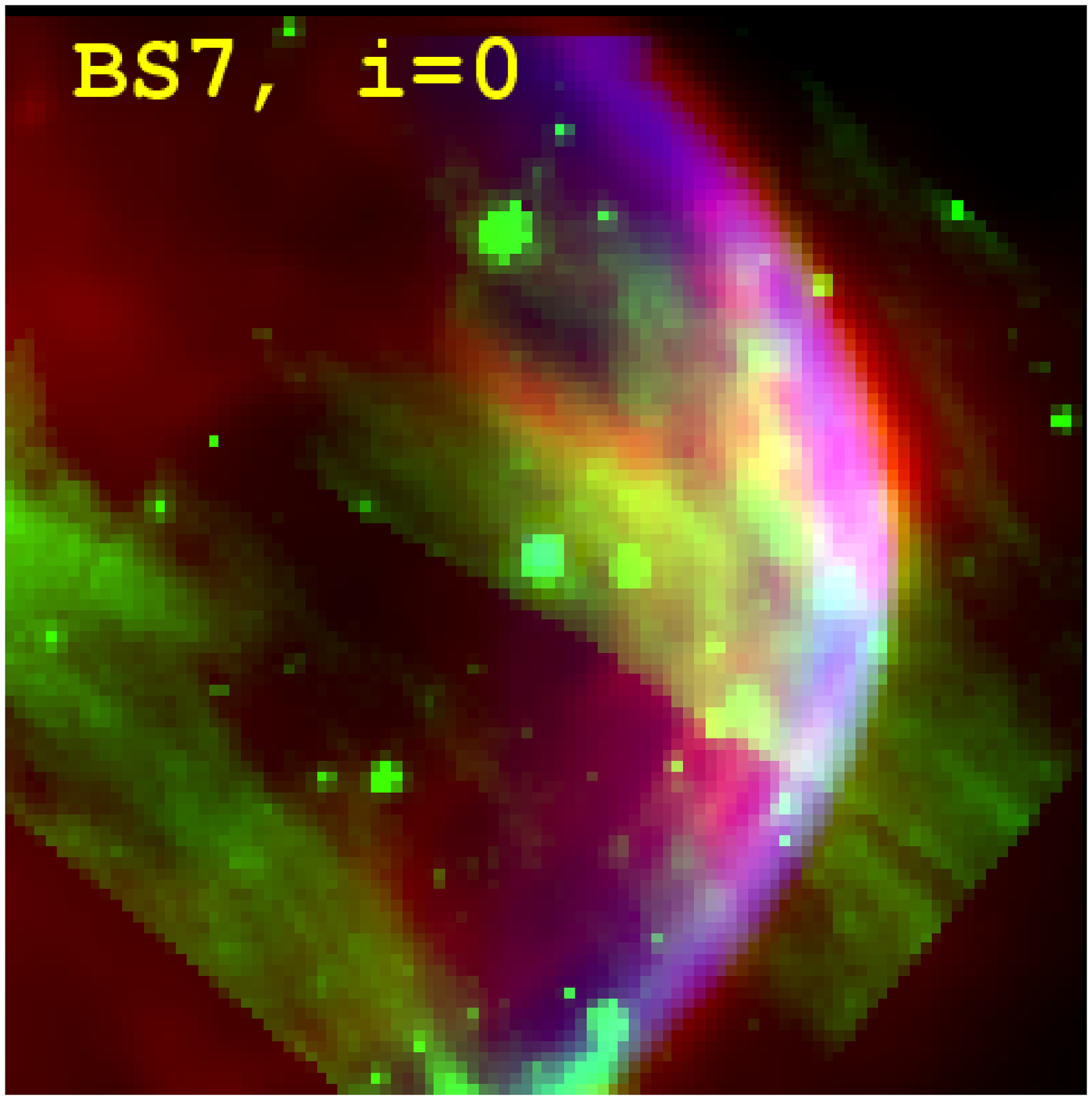}
\caption{Images of selected bowshocks with [24] in red, [8.0] in green, and a simulated
bowshock, based on the analytical shape parameterization of \citet{wilkin96}, in blue.
The angular scales and rotations are arbitrarily adjusted to
facilitate comparison between the data and models.    
\label{sim} }
\end{figure}

\begin{figure}
\plotone{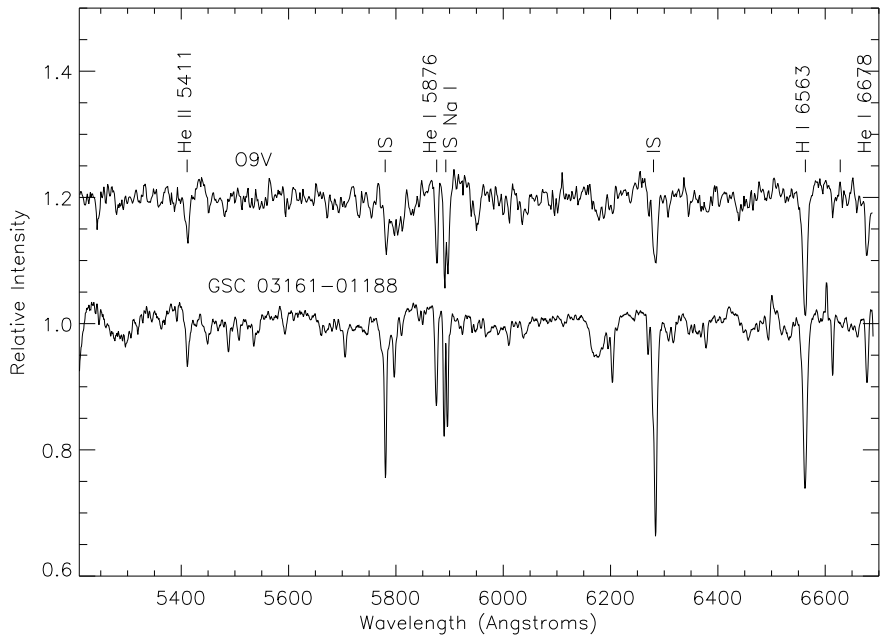}
\caption{Standard O9V spectrum from \citet{jacoby} (top) and WIRO
spectrum of Star 1 (GSC03161-01188) (bottom). The relative strengths of the \ion{H}{1} 6563 \AA,
\ion{He}{1} 6678 \AA\ and \ion{He}{2} 5411 \AA\ lines lead us to classify GSC03161-01188 as
a O9V. IS denotes interstellar bands.
\label{spec1} }
\end{figure}

\begin{figure}
\plotone{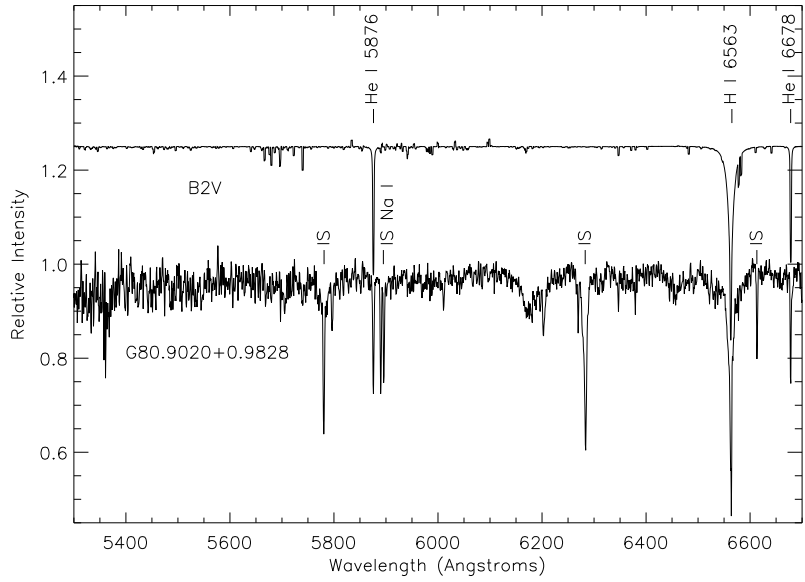}
\caption{A model atmosphere B2V equivalent spectrum from \citet{lanz} (top) smoothed to the resolution of the data and WIRO
spectrum of Star 2 (bottom). The relative strengths of the
\ion{H}{1} 6563 \AA and the \ion{He}{1} 6678 \AA\ and \ion{He}{1} 
5876 \AA\ lines lead us to classify
this star as B2V, plus or minus two spectral subclasses. 
IS denotes interstellar bands.
\label{spec2} }
\end{figure}

\begin{figure}
\plotone{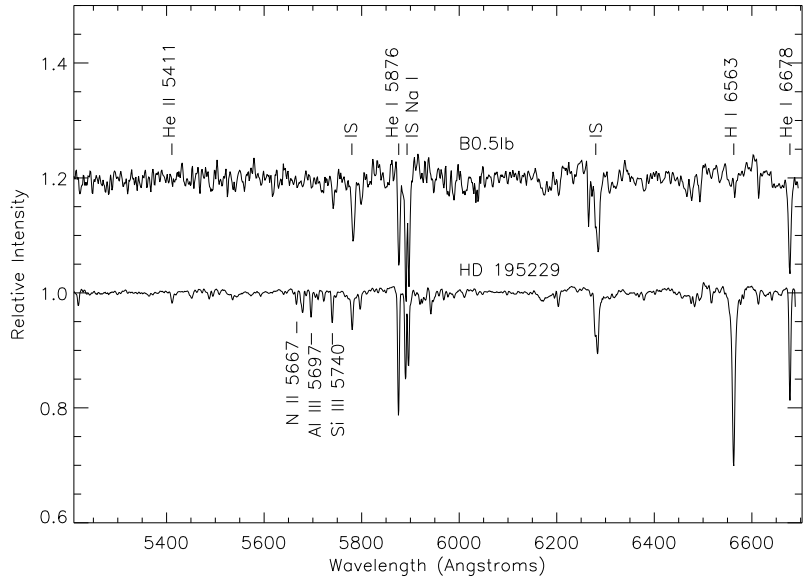}
\caption{Standard B0.2III spectrum from \citet{jacoby} (top) and WIRO
spectrum of Star 3 (HD195229) (bottom). The WIYN spectrum of Star 3 
in Figure~\ref{spec4} was used to determine this star's spectral type.
\label{spec3} }
\end{figure}

\begin{figure}
\plotone{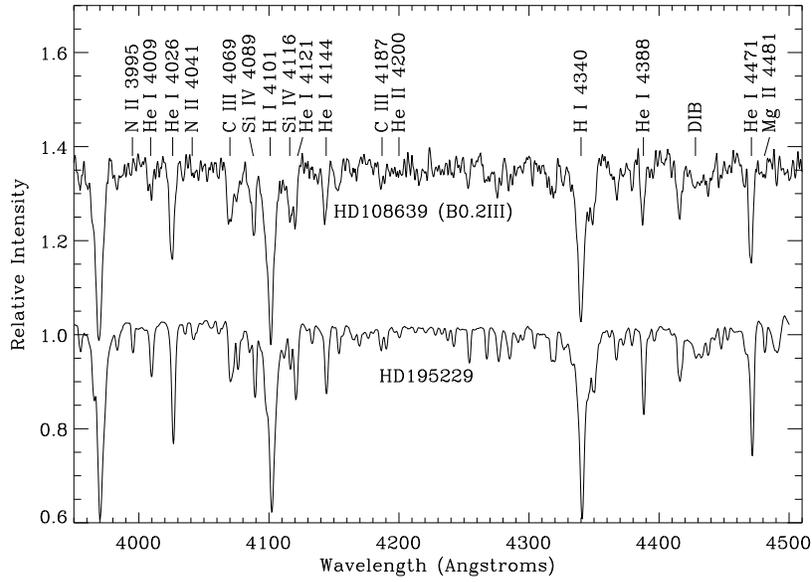}
\caption{Standard B0.2III spectrum of HD108639 \citep{walborn} and
WIYN spectrum of Star 3 (HD195229) (bottom). The strengths of the \ion{Si}{4} 4089,
 \ion{Si}{4} 4116 \AA\ and \ion{C}{3} 4069 \AA\ lines identify the star as B0.2III.
\label{spec4} }
\end{figure}

\begin{figure}
\plotone{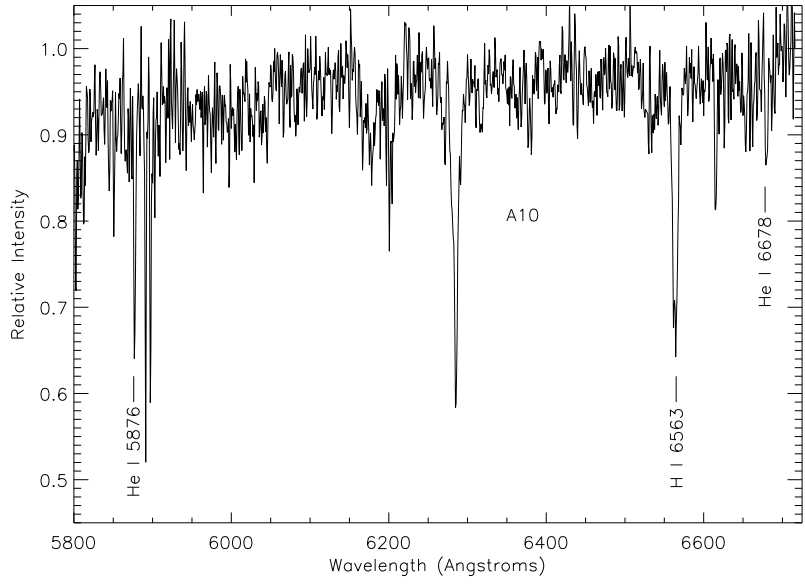}
\caption{WIRO spectrum of Star 5, also known as A10 in the notation of
\citet{comeron02}.  The spectrum is consistent with a late O star, although we estimate
the uncertainty to be several spectral subclasses owing to the low signal-to-noise ratio of the spectrum.  
\label{specA10} }
\end{figure}

\noindent $R_0$ is readily observed from the bowshock images, assuming
a 1.6 kpc distance to Cygnus~X and assuming $cos i \simeq 1$.  We
assume a minumum star--ISM speed of $V_a$=30 \kms\ for runaway stars
\citep{gies86}, and we adopt the stellar wind speeds of
\citet{mokiem07} which, for our sample of O5 -- B0 stars, range from
3000 \kms\ to $\sim$800 \kms\ ($V_{w,8}$=3 to $<$0.8).  The ambient number
density, $n_{a,3}$, is difficult to measure directly, but may be estimated
from  $n_{s,3}$, the post-shock density within the luminous bowshock
nebulae.  For strong, highly supersonic shocks,  $n_{s,3} \simeq 4  n_{a,3}$ 
\citep{landau}.  Without knowing the Mach number of each individual case, 
we adopt  $n_{s,3} = 2 n_{a,3}$, and consider the uncertainties to be factors of 
2--4.  We estimate $n_{s,3}$ by fitting the bowshock spectral
energy distributions in Figure~\ref{BSsed} with the dust/PAH models of
\citet{draineli07} which give the dust emissivity per H nucleon as a
function of the radiation field intensity, parameterized as a
multiple, $U$, times the mean interstellar radiation field of
$U_{ISRF}$=2.3$\times10^{-2}$ erg s$^{-1}$ cm$^{-2}$ \citep{mathis83}.
We make the motivated assumption that the bowshock luminosity stems
from reprocessed photons from the central star rather than thermalized
mechanical energy from the stellar wind \citep{benjamin}.  The
appropriate choice of model radiation field intensity, $U$, is then
the ratio of the flux from the central star at the standoff distance
over the mean interstellar field, estimated as,

\begin{equation}
U = {{R_*^2 \sigma T_*^4}\over{R_0^2}} ~U_{ISRF}^{-1} ~,
\end{equation}   

\noindent where $R_*$ is the stellar radius, $T_*$ the effective stellar temperature,
and $\sigma$ the usual Stefan-Botlzmann constant.  The postshock number density can then be estimated as 
\begin{equation}
n_{s,3} = 10^{-3} {{N} \over{V}} ~ ,
\end{equation}

\noindent where V is the bowshock volume which we estimate by
approximating the bowshocks as hollow cones at a known distance, $D$,
and $N=L_{obs}/L_{0}$, the ratio of the observed luminosity at a given
bandpass to the theoretical luminosity per nucleon in the
\citet{draineli07} models.  Effectively, $n_{s,3}$ is found
empirically because it is the scale factor by which we multiply the
\citet{draineli07} models to match the observed SED.  Because $V\propto
D^3$ and the inferred luminosity $L_{obs}\propto D^2$, our estimate
for $n_{s,3}$ varies as $D^{-1}$, inversely with the adopted distance.

\citet{draineli07} parameterize the model spectral energy
distributions in terms of a maximum and minimum radiation field
indicent on the dust, $U_{max}$ and $U_{min}$.  While dust in the
typical ISM is exposed to a range of intensities, we expect that
material in the bowshocks is irradiated by a single intensity dictated
by the stellar luminosity and standoff distance. Therefore, we adopt
the models with $U_{max}=U_{min}\equiv U$, where $U$ is chosen
according to Equation~7.  Figure~13 of \citet{draineli07} shows that
the flux ratio [24]/[70] is sensitive to $U$ over the range
$1<U<10^7$, and we make use of this sensitivity in fitting the
bowshock SEDs below.  We adopt the models appropriate to Milky Way
dust and leave the PAH fraction, $q_{PAH}$, as a free parameter.
Larger values of $q_{PAH}$ raise the flux in the IRAC bandpasses where
PAH emission is strong, but have no effect on the SED at [24] and
[70].

Dotted lines in Figure~\ref{BSsed} show illustrative dust models that
provide reasonable fits to the mid- and far-IR data.  The slope of the
SED between [24] and [70] allows us to guess the appropriate value for
$U$ {\it without} any prior knoweldge of the the stellar luminosity or
$R_0$.  Interestingly, the values for $U$ in each case calculated from
Equation~7 are generally within factors of 2--3 of the initial guess.
This correspondance serves as a kind of corroborating evidence that
the bowshocks are, in fact, powered by the radiant energy from their
central stars.  The fits of the models to the data in
Figure~\ref{BSsed} are done by eye to maximize the agreement at [24]
and [70] by adjusting $n_{s,3}$, and then the parameter $q_{PAH}$ is
varied to achieve a best fit at the IRAC bands.  The models provide
reasonably good fits to the data for objects 3, 5, \& 8.  In other
cases, the overall fit is poor at the IRAC bandpasses because the
model PAH features are too weak, even for the maximum model $q_{PAH}$,
or too weak, even for the minimum model $q_{PAH}$.  We take these
discrepancies as indications that PAHs may be destroyed in the
bowshocks (former case; objects 1, 4, \& 7), and that PAH excitation
may be elevated and dominate the SED in some objects (latter case;
objects 2 \& 6).  The adopted values for $U$ and $n_{s,3}$ appear in
each panel.

Armed with estimates of the ambient number densities, $n_{a,3}=1/2
n_{s,3}$, we proceed to calculate mass loss rates, $\dot M_{w,-6}$,
using the method outlined above. Again, we consider only the four
objects with the most reliable spectroscopic types where the bowshock
nature is more likely to be correct.  Table~\ref{tab.mdot} summarizes
the relevant parameters adopted for each of objects (2, 5, 7, 11).
Derived mass loss rates vary from $<0.18\times10^{-6}$ \moy\ for the
early B star (Star 2) to $\sim1\times10^{-6}$ \moy\ for the O9V star
(Star 5), to $\sim3\times10^{-6}$ \moy\ for the O5V star (A37). The
O4If star BD+43{\degr}3654 stands out, having an implied mass loss
rate of 160$\times10^{-6}$ \moy.  This large value results from the
rather large standoff distance, $R_0=1.4$ pc, implied by the $MSX$
infrared image from \citet{comeron07}.  We note that we have no direct
measure of the ambient ISM density for this object, so we have assumed
the average value from the other four early type bowshock stars.
Results for BD+43{\degr}3654 should be regarded as particularly
uncertain.   These results are broadly consistent with the mean $\dot M$
values for mid- to late-O stars estimated by \citet{lamers,
fullerton06} but a factor of several or more larger than the more
recent compilation of \citet{mokiem07}.
 
We consider the uncertainties on mass loss rates to be factors of
2--4, dominated by the uncertainties on $n_{a,3}$, which are of the
same magnitude.  The other parameters that appear in Equation~6 are
likely known to $\sim$30\%, so that uncertainties in the standoff
distance, which enter as $R_0^2$ and $V_a^2$, do not have a major
impact on the result.  
We present the calculation of mass loss rates in
this section as an outline of a novel method without claiming that
this approach is in any sense superior to other well-established
techniques.

\section{Conclusion}

We have identified ten candidate bowshock nebulae and their central
stars in the region surrounding Cyg~OB2 on the basis of mid-IR
morphologies.  We provide the first spectral types and radial
velocities for several stars (objects 1, 2, 5, \& 8) from this survey,
showing that they are late-O or early B dwarfs.  We find HD195229 
to be a likely runaway on the basis of its space velocity calculated from
both radial velocity measurements and Hipparcos proper motions, but the
velocity vectors implied by the proper motion data and bowshock orientation are nearly orthogonal.   We
measure the radial velocity of the suspected runaway O4If star
BD+43{\degr}3654 to be $V_{helio}=-66\pm$4 \kms, supporting its
runaway status.  Based on the morphologies, spectral types, and
spectral energy distributions, we identify three objects (6, 8, 9) as
probable B-star young stellar objects with circumstellar material.
Objects 2, 5, and 7 are probable bowshocks based on similarity to the
theoretical shape. These stars, however, have modest radial
velocities, consistent with those of other Cyg~OB2 members, indicating
that their motions must be predominantly tangential to the line of sight.
Without proper motion data, we are not able to confirm a runaway
status or suggest an origin for these stars.  The physical nature of
the nebulae around stars 1, 3, 4, and 10 remains ambiguous with the
current data.  These nebulae may be partial shells, partial bubbles,
or rims of nearby clouds illuminated by the early-type stars.  As part
of our analysis, we have also proposed a novel method for determining
the mass loss rates for runaway OB stars using bowshock physics that
yield reasonable, if slightly larger values for $\dot M$ compared to
more traditional methods.

\acknowledgments

We would like to thank the time allocation committees for the WIYN and
WIRO observatories for graciously allowing us the observing time that
made this project possible. We also thank the National Science
Foundation for support provided through the Research Experience for
Undergraduates (REU) grant AST 03-53760. We also thank Heather Choi
for assisting with the observing run at WIRO.  Bob Benjamin and Matt
Povich provided helpful feedback on early drafts of this manuscript.
Chris Brunt graciously allowed us early access to the CO survey of
Cygnus~X.  Advice from an astute referee strengthened the analysis
herein.

This publication makes use of data products from the Two Micron All
Sky Survey, which is a joint project of the University of
Massachusetts and the Infrared Processing and Analysis
Center/California Institute of Technology, funded by the National
Aeronautics and Space Administration and the National Science
Foundation.

{\it Facilities:} \facility{WIRO ()}, \facility{Spitzer () }, \facility{WIYN () }

{}

\clearpage

\begin{deluxetable}{rrrrrrrrrrrrcccc}
\rotate
\tabletypesize{\scriptsize}
\setlength{\tabcolsep}{0.02in}
\tablewidth{9.6in}
\tablecaption{Stellar Sources Associated with Bowshock Candidates \label{star.tab}}
\tablehead{
\colhead{\#} &
\colhead{R.A.} &
\colhead{Dec} &
\colhead{$\ell$} &
\colhead{$b$} &
\colhead{J} &
\colhead{H} &
\colhead{K} &
\colhead{[3.6]} &
\colhead{[4.5]} &
\colhead{[5.8]} &
\colhead{[8.0]} &
\colhead{[24]} &
\colhead{S.T.} &
\colhead{V$_{helio}$} &
\colhead{Name}  \\
\colhead{} &
\colhead{(J2000)} &
\colhead{(J2000)} &
\colhead{} &
\colhead{} &
\colhead{(mag)} &
\colhead{(mag)} &
\colhead{(mag)} &
\colhead{(mJy)} &
\colhead{(mJy)} &
\colhead{(mJy)} &
\colhead{(mJy)} &
\colhead{(mJy)} &
\colhead{} &
\colhead{(km~s$^{-1}$)} &
\colhead{}}
\startdata
1 & 20:34:28.9 & +41:56:17.0 & 80.8621 &  0.9749 &   9.02(0.02)  &   8.45(0.01)  &  8.21(0.01) & 138(2) & 91(2)  & 79(4)  & 47(4)   & \nodata & O9V\tablenotemark{a}                  & -17$\pm$6  &  GSC03161-01188  \\
2 & 20:34:34.5 & +41:58:29.3 & 80.9020 &  0.9828 &  11.35(0.02)  &  10.78(0.02)  & 10.51(0.02) & 32(2)  &  35(3) & 74(6)  & 161(8)  & \nodata & B1V--B3V\tablenotemark{a}             & -12$\pm$15 &  G80.9020+0.9828   \\
3 & 20:28:30.2 & +42:00:35.2 & 80.2632 &  1.9137 &   7.34(0.02)  &   7.36(0.01)  &  7.32(0.02) & 287(8) & 204(7) & 126(6) & 64(4)   & \nodata & B0.2III\tablenotemark{a}              & -3$\pm$2   &  HD195299	        \\     
4 & 20:28:39.4 & +40:56:51.0 & 79.4172 &  1.2703 &  14.25(0.03)  &  12.31(0.02)  & 11.32(0.01) & 19(2)  & 14(2)  & 10(3)  & 17(2)   & \nodata & B2V--B3V\tablenotemark{b}             & \nodata    &  G79.4172+1.2703  \\
5 & 20:34:55.1 & +40:34:44.0 & 79.8224 &  0.0959 &   9.54(0.02)  &   8.70(0.02)  &  8.25(0.02) & 185(2) & 127(3) & 96(4)  & 131(6)  & \nodata & O9V\tablenotemark{a} \nodata          & -10$\pm$10   &  A10\tablenotemark{c}  \\
6 & 20:36:13.3 & +41:34:26.1 & 80.7657 &  0.4966 &  16.04(0.08)  &  13.90(0.06)  & 12.76(0.03) & 23(2)  & 35(4)  & 68(4)  & 133(6)  & 1245(24)&  B4V--B6V\tablenotemark{b}\nodata     & \nodata    &  G80.7657+0.4966 (YSO?)   \\
7 & 20:36:04.4 & +40:56:13.0 & 80.2400 &  0.1354 &   8.56(0.03)  &   7.97(0.01)  &  7.68(0.02) & 376(3) & 141(2) & 135(7) & 54(3)   & \nodata &  O5V\tablenotemark{a}                 & \nodata    &  A37\tablenotemark{c} \\
8 & 20:20:11.6 & +39:45:30.1 & 77.5168 &  1.9047 &  10.5 (0.02)  &  10.00(0.02)  &  9.71(0.01) & 34(2) & 26(2) &  59(4)   & 157(8)  & 4600(80)& B1V--B3V\tablenotemark{a}             & 2$\pm$4    &  G77.5168+1.9047 (YSO?)      \\     
9 & 20:25:43.9 & +38:11:13.2 & 76.8437 &  0.1231 & \nodata       & \nodata       &\nodata      & 8(2)  & 20(3) &  41(2)   & 63(3)   & 677(12) & B??\tablenotemark{b}                  & \nodata    & G76.8437-0.1231 (YSO?)	  \\
10& 20:29:22.1 & +37:55:44.3 & 77.0511 & -0.6092 &  10.33(0.02)  &   9.63(0.02) &  9.27(0.01)  & 74(7) & 49(9) &  50(3)   & 81(5)   & \nodata &  B1V--B2V\tablenotemark{b}            & \nodata    & G77.0511-0.6092        \\
11& 20:33:36.1 & +43:59:07.4 & 82.4100 &  2.3254 &   6.64(0.02)  &   6.19(0.02) &  5.97(0.02)  & \nodata &\nodata &\nodata&\nodata  & \nodata & O4If\tablenotemark{a}                 & -66$\pm$4  & BD+43{\degr}3654  \\
\enddata
\tablenotetext{a}{Spectral type based on spectroscopic identification. }
\tablenotetext{b}{Spectral type adopted by assuming a main-sequence star consistent with the near-IR SED if at the nominal 1.6 kpc distance of Cygnus~X. }
\tablenotetext{c}{\citet{comeron02}}
\end{deluxetable}

\clearpage

\begin{deluxetable}{rrrrrrrrrrrrr}
\tabletypesize{\scriptsize}
\tablecaption{Photometry of Bowshock Candidates\label{phot.tab}}
\tablewidth{0pt}
\tablehead{
\colhead{\#} &
\colhead{[3.6]} &
\colhead{$\sigma_{[3.6]}$} &
\colhead{[4.5]} &
\colhead{$\sigma_{[4.5]}$} &
\colhead{[5.8]} &
\colhead{$\sigma_{[5.8]}$} &
\colhead{[8.0]} &
\colhead{$\sigma_{[8.0]}$} &
\colhead{[24]} &
\colhead{$\sigma_{[24]}$} &
\colhead{[70]} &
\colhead{$\sigma_{[70]}$}  \\
\colhead{} &
\colhead{(mJy)} &
\colhead{(mJy)} &
\colhead{(mJy)} &
\colhead{(mJy)} &
\colhead{(mJy)} &
\colhead{(mJy)} &
\colhead{(mJy)} &
\colhead{(mJy)} &
\colhead{(mJy)} &
\colhead{(mJy)} &
\colhead{(mJy)} &
\colhead{(mJy)}}
\startdata
1  & $<$36 & \nodata  & $<$ 35 & \nodata &  $<$ 73  & \nodata & $<$ 122 & \nodata  &  5100 & 1020  &  1220 &  250 \\
2  &    78 &      16  &     88 &      18 &     328  &  65     &     795 &  159     &   539 &  108  & 10300 & 2060 \\
3  &    20 &      10  &     19 &      10 &   $<$14  & \nodata & $<$  20 & \nodata  &  1370 &  274  &   659 &  132 \\
4  & $<$ 5 & \nodata  & $<$  6 & \nodata &  $<$  5  & \nodata & $<$  12 & \nodata  &  1900 &  380  &  2900 &  580 \\
5  &    78 &      16  &     92 &      18 &     158  &  32     &    1288 &  257     & 65300 &13000  & 73000 &14600 \\
6  &   160 &      32  &    100 &      20 &    1480  & 296     &    3900 &  780     &\nodata\tablenotemark{a}&\nodata& 10970 & 2190 \\
7  & $<$26 & \nodata  & $<$ 21 & \nodata &  $<$ 23  & \nodata & $<$  67 & \nodata  & 47220 & 9440  & 35300 & 7060 \\
8  &    71 &      14  &     88 &      18 &     214  &  43     &     570 &  114     & 2394  &  478  & 15400 & 3080 \\
9  & $<$ 4 & \nodata  & $<$  4 & \nodata &      46  &  10     &     118 &   24     &  276  &   57  &  3800 &  760 \\
10 &    21 &       4  &     24 &       5 &      32 &    7     &     100 &   20     & 1878  &  360  &   182 &   36 \\
\enddata
\tablecomments{Uncertainties are estimated at 20\% in all cases except nondetections where 1$\sigma$ upper limits are given.  }
\tablenotetext{a}{High and variable background precludes a reliable measurement at [24] for this source.  }
\end{deluxetable}

\begin{deluxetable}{rccccc}
\tabletypesize{\scriptsize}
\tablecaption{Calculated Bowshock Parameters for Most Reliable Objects \label{tab.params}}
\tablewidth{0pt}
\tablehead{
\colhead{\#} &
\colhead{S.T.} &
\colhead{$R_0~cos~i$} &
\colhead{$\dot M_{w,-6}$} &
\colhead{$V_{w,8}$} &
\colhead{$V_a n_a^{1/2}$ }   \\
\colhead{} &
\colhead{} &
\colhead{(pc)} &
\colhead{($10^{-6}$ M$_\odot$ yr$^{-1}$) } &
\colhead{($10^8$ cm s$^{-1}$)} &
\colhead{(km s$^{-1}$ (10$^3$ cm$^{-3}$)$^{1 \over 2}$}) }
\startdata
2   & B1V--B3V\tablenotemark{c}   &  0.07 & 0.0003 - 0.001  & $<$0.8      & $<$0.68      \\
5   & O9V\tablenotemark{a}        &  0.19 &  0.0003 - 0.1   & 0.8 -- 1.5  & 0.12 -- 3   \\
7   & O5V\tablenotemark{a}        &  0.53 &  0.2 -- 0.9     & 2.8 --  3.2 &  2 -- 5   \\
11   & O4If\tablenotemark{a}       &  1.4\tablenotemark{b}  & 8             &  2.2          &  11  \\
\enddata
\tablenotetext{a}{Based on spectroscopic identification.}
\tablenotetext{b}{Based on the images of \citet{comeron07} and adopted 1.6 kpc distance.}
\end{deluxetable}

\begin{deluxetable}{rccc}
\tabletypesize{\scriptsize}
\tablecaption{Derived Mass Loss Rates for Most Reliable Objects\label{tab.mdot}}
\tablewidth{0pt}
\tablehead{
\colhead{\#} &
\colhead{S.T.} &
\colhead{$n_{s,3}$} &
\colhead{$\dot M_{w,-6}$}    \\
\colhead{} &
\colhead{} &
\colhead{(10$^3$ cm$^{-3}$)}  &
\colhead{(10$^{-6}$ M$_\odot$ yr$^{-1}$}) }
\startdata
2   & B1V--B3V   &  0.10 & $<$0.18    \\
5   & O9V        &  0.1  & 0.75 - 1.3       \\
7   & O5V        &  0.10 &  2.5 -- 3           \\
11   & O4If       &  0.10\tablenotemark{a} & 160            \\
\enddata
\tablenotetext{a}{Not measured; value adopted from the average of othe ther stars.}
\end{deluxetable}

\clearpage

\clearpage

\clearpage

\clearpage

\clearpage

\clearpage

\clearpage

\clearpage

\clearpage

\clearpage

\clearpage

\clearpage

\clearpage

\clearpage

\clearpage

\clearpage

\clearpage

\clearpage

\clearpage

\clearpage

\begin{figure}
\plotone{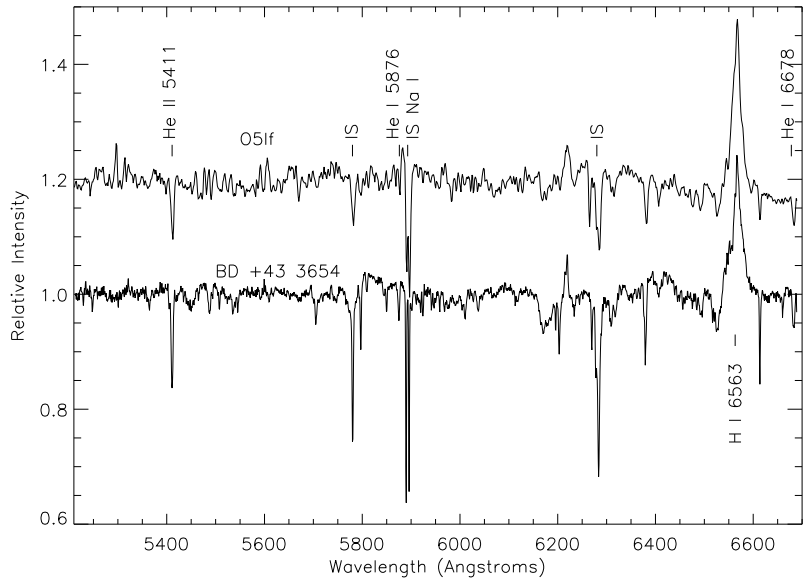}
\caption{WIRO spectrum of BD+43\degr3654. Comparison with a standard
O5If from \citet{jacoby} (the earliest O supergiant in this library)
agrees with the O4If classification given by \citet{comeron07}.
\label{spec5} }
\end{figure}

\clearpage

\end{document}